\documentclass[floats,prd,notitlepage,preprintnumbers,longbibliography,nofootinbib]{revtex4-1}
\usepackage[colorlinks=true,urlcolor=blue,linkcolor=blue,citecolor=blue]{hyperref}
\usepackage{slashed,color,amsmath,amssymb,mathrsfs}
\usepackage{amsfonts,epsfig,amssymb}
\usepackage{graphicx}
\usepackage{hyperref}
\usepackage{longtable}
\usepackage{array}
\usepackage{accents}
\usepackage{tensor}
\usepackage{footmisc}
\usepackage{longtable}
\usepackage{booktabs}
\usepackage{multirow}
\usepackage{bm}
\usepackage{subfigure}

\allowdisplaybreaks

\begin{document}

\title{Study of $X(6900)$ with unitarized coupled channel scattering amplitudes  }
\author{Shi-Qing Kuang$^{1,2}$}
\author{Qi Zhou$^{1,2}$}
\author{Di Guo$^{1,2}$}
\author{Qin-He Yang$^{1,2}$}
\author{Ling-Yun Dai$^{1,2}$}
\email{dailingyun@hnu.edu.cn}
\affiliation{$^{1}$ School of Physics and Electronics, Hunan University, Changsha 410082, China}
\affiliation{$^{2}$ Hunan Provincial Key Laboratory of High-Energy Scale Physics and Applications, Hunan University, Changsha 410082, China}
\date{\today}
\begin{abstract}
In this paper, we study the resonant state $X(6900)$. The scattering amplitudes of coupled channels, $J/\psi J/\psi$-$J/\psi \psi(2S)$-$J/\psi \psi(3770)$, are constructed with the interaction of four vector mesons described by effective Lagrangians. The amplitudes are calculated up to one loop, decomposed by partial wave projection, and unitarized by Pad$\acute{e}$ approximation. These amplitudes are fitted to the latest experimental data sets of di-$J/\psi$ and $J/\psi \psi(2S)$ invariant mass spectra of LHCb, CMS, and ATLAS. High-quality solutions are obtained. With these partial wave amplitudes, we extract the pole parameters of the $X(6900)$. Its quantum number is likely to be $0^{++}$. According to the pole counting rule as well as analysis of the phase shifts of the partial waves, it supports our previous conclusion that the $X(6900)$ prefers to be a compact tetra-quark.  
\end{abstract}

\maketitle

\section{Introduction}
\label{Sec:I}
Since the 1960s, quark model \cite{Gell-Mann:1964ewy,Zweig:1964ruk,Zweig:1964jf} has turned out to be a successful classification scheme for hadrons, that is, a meson is composed of a pair of quark and anti-quark, and a baryon/anti-baryon is composed of three quarks/anti-quarks. Hundreds of hadrons listed in the particle data group (PDG) \cite{ParticleDataGroup:2020ssz} can be formed as such inner structure. 
However, until now, there is no fundamental principle to rule out other inner structures of hadrons, such as hadronic molecules, quark-gluon hybrids, glueballs, and multi-quark states. Searching for exotic states remains a keen interest of the physics community. In 2003, a tetraquark candidate, $X(3872)$, was discovered by Belle \cite{Belle:2003nnu} and other collaborations \cite{BaBar:2004oro,CDF:2003cab,D0:2004zmu}. In 2013, BESIII and Belle discovered the $Z_c^+$(3900)\cite{Ablikim:2013mio,Liu:2013dau} in the $J/\psi\pi^+$ invariant mass spectrum, implying the $\bar{c}c\bar{d}u$ component. In 2015, LHCb discovered two $P_c$ states  \cite{Aaij:2015tga}  and later they are found to be \lq splitted' into three such states \cite{Aaij:2019vzc}, $P^+_c$(4312), $P^+_c$(4440), $P^+_c$(4457), with about nine more times decay events collected.  The $P_c$s are observed in the $J/\psi p$ invariant mass spectra, and hence they should contain at least five quarks, $\bar{c}cuud$. In 2021, 
LHCb discovered the $T_{cc}^+$ in $D^0D^0\pi^+$ invariant mass spectrum \cite{LHCb:2021vvq,LHCb:2021auc}. This further grasps the attention of theorists as it is very likely to be evidence for multi-quark states, $cc\bar{u}\bar{d}$. 
All these exotic hadrons observed so far contain at most two heavy (charm or bottom) quarks/anti-quarks, but as many models predicted, there should also be multi-quark states composed of three or more heavy quarks/anti-quarks \cite{Iwasaki:1976cn,Berezhnoy:2011xn,Wu:2016vtq,Liu:2019zuc,Anwar:2017toa}.     
This is confirmed by the very recent experiment of LHCb collaboration \cite{LHCb:2020bwg}. 
With the datasets of proton-proton collision in the center-of-mass energies $\sqrt{s}=7,8$ and 13 TeV collected by the LHCb detector, the invariant mass spectrum of $J/\psi J/\psi$ was measured in the energy range of [6.2-7.4]~Gev/$c^2$, and a narrow structure was found near 6.9Gev/$c^2$, labeled as $X(6900)$, with the signal statistical significance being larger than 5~$\sigma$. Its mass and width are given in two ways: 
On the one hand, they assume that the non-resonant single-parton scattering (NRSPS) continuum is not disturbed, and Breit-Wigner forms fit the structure. Then the mass and width are determined to be 
\begin{eqnarray}
M[X(6900)]&=&6905\pm 11\pm7 ~{\rm MeV}/{c^2}\,, \nonumber\\
\Gamma[X(6900)]&=&80\pm 19\pm33 ~{\rm MeV}/{c^2}\,. \nonumber
\end{eqnarray}
%with the uncertainties being statistical and systematic ones in order. 
On the other hand, once the contribution of the NRSPS continuum is taken into account, one has  
\begin{eqnarray} 
M[X(6900)]&=&6886\pm 11\pm11 ~{\rm MeV}/{c^2}\,, \nonumber\\
\Gamma[X(6900)]&=&168\pm 33\pm69 ~{\rm MeV}/{c^2}\,. \nonumber
\end{eqnarray}  
Recently, ATLAS and CMS presented their measurements on $J/\psi J/\psi$ invariant mass spectra, too. They confirmed the existence of the $X(6900)$. For CMS, the mass and width of the $X(6900)$ are given as \cite{Zhang:2022toq}
\begin{eqnarray} 
M[X(6900)]&=&6.87\pm 0.03^{+0.06}_{-0.01}~{\rm GeV}\,, \nonumber\\
\Gamma[X(6900)]&=&0.12\pm0.04^{+0.03}_{-0.01}~{\rm GeV}\,. \nonumber
%M[X(6600)]&=&(6.62\pm 0.03^{+0.02}_{-0.01})GeV\,, \nonumber\\ 
%\Gamma[X(6600)]&=&(0.31\pm0.09^{+0.06}_{-0.11})GeV \,.\nonumber
\end{eqnarray}
and ATLAS measures the mass and width of the $X(6900)$ as \cite{Xu:2022rnl}
\begin{eqnarray} 
 M[X(6900)]&=&6927\pm 9\pm 5~{\rm MeV}\,, \nonumber\\
\Gamma[X(6900)]&=&122\pm 22\pm 19~{\rm MeV}\,.  \nonumber
%M[X(6600)]&=&(6552\pm 10\pm 12)MeV\,, \nonumber\\
%\Gamma[X(6600)]&=&(124\pm 29\pm 34)MeV\,. \nonumber
\end{eqnarray} 
%%%%%
Not limited to confirming the $X(6900)$, they also find a new fully heavy quark state, $X(6600)$. 
These measurements certainly should be included in the analysis. 

One would notice that the $X(6900)$ is discovered in the di-$J/\psi$ invariant mass spectrum, which is also suggested by earlier theoretical prediction Ref.\cite{Chen:2016jxd}.
Hence, it is not a stretch to infer that this state is composed of at least two charm quarks and two charm anti-quarks, resulting in a cornucopia of models trying to classify its origin and search for similar states, see e.g., Refs.~\cite{Dong:2020nwy,Wang:2020wrp,Gong:2020bmg,Cao:2020gul,Guo:2020pvt,Liang:2021fzr,Wang:2020gmd,Ke:2021iyh,Chen:2020xwe,Wang:2022jmb,Zhang:2022qtp,Wang:2022yes,Dong:2022sef,Wang:2022xja}. 
Nonetheless, before discussing its inner structure, a natural and fundamental problem is determining the mass, width, and quantum number of the $X(6900)$. 
%%%%
Among the theoretical research, Ref.~\cite{Dong:2020nwy} unitarize the amplitudes and fit them to the invariant mass spectrum. They extract the mass and width of the $X(6900)$ as $6818^{+28}_{-32}-i142^{+14}_{-10}$~MeV in the third Riemann sheet in their two-channel case, with the quantum number either to be $0^{++}$ or $2^{++}$, while in the triple-channel case they do not find such a pole. They also claim a near-threshold resonance named $X(6200)$.
In Ref.~\cite{Cao:2020gul}, they discuss the nature of the $X(6900)$ and conclude that it could very well be confining states or molecular states, but it is impossible to distinguish them at present. In Ref.~\cite{Chen:2020xwe}, they find that the wide peaks between 6200 and 6800~MeV are caused by contributions of S-waves, with the quantum number to be either $0^{++}$ or $2^{++}$. While the $X(6900)$ can be regarded as a P-wave resonance, with the quantum number to be either $0^{-+}$ or $1^{-+}$. Nevertheless, these works are not originated from partial wave decomposition, and they assign a quantum number to the resonance. Correspondingly, only a sole partial wave is taken into account when fitting the data. 

%%%%
In our earlier work \cite{Zhou:2022xpd}, we perform an amplitude analysis to extract the pole parameters (mass, width, and residues). Nonetheless, the previous work is only about the LCHb's data, and we need to update it to include the new measurements from ATLAS and CMS. The strategy is similar to before: The amplitudes are calculated up to one loop with effective Lagrangians, and partial wave projections are applied.
Pad\'e approximation is used to perform unitarization, and the scattering amplitudes are constructed. By fitting to the experimental data, the unknown couplings of the effective Lagrangians are fixed, and thus one can obtain the pole information of the resonance at last. 
%%%%
With the pole information, one can study the nature of the $X(6900)$ according to the pole counting rule  \cite{Morgan:1992ge,Dai:2011bs,Dai:2012kf}, which is helpful to distinguish molecule or Breit-Wigner type origins. The conclusion can be further tested by extracting the phase shifts of the scattering amplitudes. As has been recognized in the last few decades,  phase shift is one of the most critical inputs for S-matrix methods such as dispersion relation \cite{Ishida:1995xx,Ishida:1997wn}. Obtaining exact phase shifts would help to confirm the existence and also give clues for the nature of the resonance \footnote{Typical examples can be found in research for light scalars \cite{Xiao:2000kx,Zhou:2004ms,Caprini:2005zr,Descotes-Genon:2006sdr}.}. Therefore, we will also extract the phase shifts of the $J/\psi J/\psi$ partial wave scattering amplitudes.

In the following part of the paper, in Sec.~\ref{Sec:II}, we build the scattering amplitude and perform its partial wave decomposition.
In Sec.~\ref{Sec:III}, we fit our amplitudes to the invariant mass spectrum of $J/\psi$-pairs,  and determine the mass and width of the $X(6900)$ as well. Finally, we give the conclusions in Sec.~\ref{Sec:IV}.

\section{Formalism} \label{Sec:II}
\subsection{Effective Lagrangians and scattering amplitudes}
For the scattering of coupled channels $J/\psi J/\psi$-$J/\psi \psi(2S)$-$J/\psi \psi(3770)$, we construct the following effective Lagrangians\footnote{Some of the formalism has been given in Ref.~\cite{Zhou:2022xpd}, but for reader's convenience, we rewrite it and give more detials.}:
\begin{eqnarray}
    \mathcal{L}&=&c_1V_{\mu} V_{\alpha} V^\mu V^\alpha+c_2 V_{\mu} V_{\alpha} V^\mu V^{\prime\alpha}+c_3V_{\mu} V^\prime_{\alpha} V^\mu V^{\prime\alpha}+
    c_4V_{\mu} V^{\prime\mu} V_{\alpha} V^{\prime\alpha}+c_5V_{\mu} V_{\alpha} V^\mu V^{\prime\prime\alpha} \nonumber\\
    &&+c_6V_{\mu} V^{\prime\prime}_{\alpha} V^\mu V^{\prime\prime\alpha}+c_7V_{\mu} V^{\prime\prime\mu} V_{\alpha} V^{\prime\prime\alpha}+c_8V_{\mu} V^\prime_{\alpha} V^\mu V^{\prime\prime\alpha}+c_9V_{\mu} V^{\prime\mu} V_{\alpha} V^{\prime\prime\alpha}\,,\label{eq:lagran}
\end{eqnarray}
where $V$, $V^\prime$, $V^{\prime\prime}$ represent for $J/\psi$, $\psi(2S)$, $\psi(3770)$, respectively. 
Note that the interactions of four heavier vector mesons, $V^{\prime,\prime\prime}$, such as $V^\prime_{\mu} V^\prime_{\alpha} V^{\prime\mu} V^{\prime\alpha}$, are ignored since the thresholds of $V^{\prime(\prime\prime)}V^{\prime(\prime\prime)}$ would be heavier than 7.2~GeV, beyond the energy region of the invariant mass spectrum we focus on. 
The $\chi_{c0}\chi_{c0}$ and $\chi_{c1}\chi_{c1}$ channels are ignored, too, as they are farther away from the structure (the peak around 6.90 GeV and the dip around 6.75 GeV) of di-$J/\psi$ spectra, compared with the $V V^\prime$ and $V V^{\prime\prime}$ channels. 
Also, the left hand cuts generated by the meson exchanges through $t$- and $u$-channels scatterings in the processes of $J/\psi J\psi \to \chi_{cJ}\chi_{cJ}$ are farther away than $J/\psi J\psi \to J\psi J\psi$, $J\psi \psi(2S)$, $J\psi \psi(3770)$. For more details, see discussions in Appendix \ref{app:chicJ}. 
For simplicity, the interactions between vectors and pseudoscalar or scalar mesons, for instance,  $VV\pi\pi$ terms,  are not taken into account as what is done in the hidden gauge symmetry formalism \cite{Bando:1984ej,Geng:2016pmf}. 
%%%%%%%%
Also, we ignore the intermediate channels of $\eta_c\eta_c$, $h_ch_c$, as they are suppressed by heavy quark spin symmetry ~\cite{Dong:2020nwy,Gong:2022hgd}, too. These effective Lagrangians are consistent with the leading order (LO) Lagrangians constructed by HQSS\cite{Casalbuoni:1996pg}. 
The heavier meson with angular momentum $J=1$ is realized in the formalism of HQSS as
\begin{eqnarray}
J &=& \frac{1+v\!\!\!/}{2}[H_\mu\gamma^\mu-\eta\gamma_5]\frac{1-v\!\!\!/}{2} \,,\nonumber\\
\bar{J} &=& \gamma^0 J^\dagger\gamma^0= \frac{1-v\!\!\!/}{2}[H_\mu^{\dagger} \gamma^\mu+\eta^\dagger\gamma^5]\frac{1+v\!\!\!/}{2} \,. \nonumber
\end{eqnarray}
Consequently, the LO interaction Lagrangian for $J/\psi J/\psi\to J/\psi J/\psi$ is 
\begin{eqnarray}
\mathcal{L}_{\rm HQSS}^{LO}&=& g_1 \langle J\bar{J}J\bar{J}\rangle\nonumber\\
&=&\langle\frac{1+v\!\!\!/}{2}{H \!\!\!\!/} \frac{1-v\!\!\!/}{2}{H \!\!\!\!/}^\dagger \frac{1+v\!\!\!/}{2}{H \!\!\!\!/}\frac{1-v\!\!\!/}{2} {H \!\!\!\!/}^\dagger\frac{1+v\!\!\!/}{2}\rangle\nonumber\\
&=&2N_C g_1 V_{\mu} V_{\alpha} V^\mu V^\alpha \, , \label{eq:L;HQSS}
\end{eqnarray}
where $N_C$ is the number of colors. 
It is the same as Eq.~(\ref{eq:lagran}) with $c_1=2N_C g_1$. The other interaction Lagrangians can be obtained in the same way.
With these effective Lagrangians, we can calculate the scattering amplitude. The Feynman diagrams up to next-to-leading order (NLO) can be seen in Ref.\cite{Zhou:2022xpd}. In the appendix \ref{app:1}, we give the concrete expression of the amplitude $T^{ij}$. The superscripts \lq $i$, $j$' are labels dor channels, with the numbers \lq 1,2,3' specified as $J/\psi J/\psi$, $ J/\psi\psi(2S)$ and $J/\psi\psi(3770)$.

\subsection{Partial wave decomposition}
To clarify the quantum number of the possible resonances that appear as the intermediate states in the scattering, we need to carry out partial wave projection for the scattering amplitudes. 
%%%%
Partial wave projection of the helicity amplitude is given as \cite{Martin1970}:
\begin{eqnarray}\label{eq:pwJ}
   T^{ij}_{\mu_1\mu_2;\mu_3\mu_4}(s,z_s) = 16\pi N_{ij} \sum_{J} (2J+1) T_{\mu_1\mu_2;\mu_3\mu_4}^{J,ij}(s)d_{\mu\mu^\prime}^J(z_s) \,.
\end{eqnarray}
Here one has $z_s=\cos\theta_s$, with $\theta_s$ the scattering angle in the center of mass frame in $s$-channel. $\mu=\mu_1-\mu_2$ and $\mu^\prime=\mu_3-\mu_4$ are the difference in the helicities of the two particles in initial or final states, respectively. $N_{ij}$ is the normalization factor caused by the property of identical particles, with $N_{11}=2$ for $J/\psi J/\psi\to J/\psi J/\psi$ amplitude, $N_{i1,1i}=\sqrt{2}$ for $J/\psi J/\psi\to J/\psi \psi(2S),J/\psi \psi(3770)$ amplitudes, and $N=1$ for others. 
$s$ is the Mandelstam variable, with $s=(p_1+p_2)^2$. $d_{\mu\mu\prime}^J$ is the standard Wigner functions defined according to rotations\cite{Martin1970}. 
The partial wave amplitudes $T_{\mu_1\mu_2;\mu_3\mu_4}^{J,ij}(s)$ can be obtained in Ref.\cite{Zhou:2022xpd}.
% \begin{eqnarray}\label{eq:pwh}
%    T_{\mu_1\mu_2;\mu_3\mu_4}^{J,ij}(s)=\frac{1}{ 32\pi N_{ij}}\int_{-1}^1  T^{ij}_{\mu_1\mu_2;\mu_3\mu_4}(s,z_s)d_{\mu\mu^\prime}^J(z_s)dz_s \,.
% \end{eqnarray}
Indeed, each polarization vector of Eq.~\eqref{eq:amp} can have three different helicities, $\{+1,0,-1\}$, as all of the quantum number of $J/\psi, \psi(2S)$, and $\psi(3770)$ are $1^{--}$. For details of the polarization vectors. See Eq.(\ref{eq:polar;vector}) in Appendix \ref{app:1}. Therefore, for each angular momentum $J$, there are as many as 81 amplitudes with different combinations of helicities.  
Nevertheless, according to discrete symmetry such as $P$ and $T$, these amplitudes can be reduced to much less independent ones. 
The conservation of the amplitudes under parity transform gives
\begin{eqnarray}\label{eq:P parity}
   T_{\mu_1\mu_2;\mu_3\mu_4}^J(s)=\frac{\eta_3\eta_4}{\eta_1\eta_2}(-1)^{s_1+s_2-s_3-s_4}T_{-\mu_1-\mu_2;-\mu_3-\mu_4}^J(s)\,,
\end{eqnarray}
where $\eta_i$ is the intrinsic parity of the $i-$th particle, and $s_i$ is the spin.
The conservation of the amplitudes under time reversal transformation gives 
\begin{eqnarray}\label{eq:T parity}
   T_{\mu_1\mu_2;\mu_3\mu_4}^J(s)=T_{\mu_3\mu_4;\mu_2\mu_1}^J(s) \,.
\end{eqnarray}
Notice that the time reversal invariance holds true only if the initial states are equal to the final states.
%%%%
According to Eq.~\eqref{eq:P parity}, the amplitudes will be reduced to 41 independent ones. With the other constraint of Eq.~\eqref{eq:T parity}, the amplitudes will be reduced again, resulting in 25 independent amplitudes in the processes of $J/\psi J/\psi\to J/\psi J/\psi, J/\psi \psi(2S)\to J/\psi \psi(2S), J/\psi \psi(3770)\to J/\psi \psi(3770)$. See discussions below.
%%%%%

In order to clarify the quantum number of the possible resonance that appears as an intermediate state in these scatterings, we need to transfer the partial wave amplitudes from $|JM\mu_1\mu_2\rangle$ to the $|JMLS\rangle$ representation. This is converted by the following operations \cite{Martin1970}
\begin{align}
    \left|JM;\mu_1\mu_2 \right\rangle=
    &\sum_{LS}(\frac{2L+1}{2J+1})^{\frac{1}{2}}\left\langle LS0\mu|J\mu\right\rangle
    \left\langle s_1s_2\mu_1,-\mu_2|S\mu\right\rangle \left|JM;LS \right\rangle \,, \label{eq:mumuLS}
\end{align}
where the Clebsch-Gordan coefficients can be found in PDG \cite{ParticleDataGroup:2020ssz}.
With Eqs.~(\ref{eq:pwJ},\ref{eq:mumuLS}), the partial wave decomposition of each helicity amplitude in the $|JMLS\rangle$ representation can be seen Eq.\eqref{eq:pwLS} in Appendix \ref{app:1}.

%%%%
Further, it is convenient to list the quantum number $J^{PC}$ of the $J/\psi J/\psi$  system to separate the partial waves.  
In di-$J/\psi$ system, the charge conjugation and parity are given by $C=(-1)^{L+S}$ and $P=(-1)^L$. A pure neutral system requires $L+S$ to be even so that one has $C=1$. Further, we only consider the lowest partial waves with $L=0,1$ and ignore all other higher partial waves. That is, five partial waves are included in total: The S-waves, $0^{++}$ and $2^{++}$; The P-waves, $0^{-+}$, $1^{-+}$, and $2^{-+}$. See Table \ref{tab:JPC}. 
%%%%%-----------------------------------
\begin{table}[htbp]
\setlength{\tabcolsep}{9mm}
\begin{tabular}{ccccc}
\hline\hline
$L$  \rule[-0.2cm]{0cm}{6mm}    & $S=0$                 &  $S=1$                                                                    &$S=2$                                                      \\ \hline
$0$  \rule[-0.2cm]{0cm}{6mm}    & $\mathbf{{0}^{{++}}~(^1S_0)}$  &  ${1}^{{+-}}$                                                             &  $\mathbf{{2}^{{++}}~(^5S_2)}$                                                 \\
$1$   \rule[-0.2cm]{0cm}{6mm}   & ${1}^{{--}}$          &$\mathbf{{{0}^{{-+}}}~(^3P_0)}\quad\mathbf{{{1}^{{-+}}}~(^3P_1)}\quad\mathbf{{{2}^{{-+}}}~(^3P_2)}$   &  ${1}^{{--}}\quad{2}^{{--}}\quad{3}^{{--}}$                  \\
$2$   \rule[-0.2cm]{0cm}{6mm}   & $\textit{{2}}^{{++}}$          &${1}^{{+-}}\quad{2}^{{+-}}\quad{3}^{{+-}}$                                 &  ${\textit{0}}^{{++}}\quad{\textit{1}}^{{++}}\quad{\textit{2}}^{{++}}\quad{\textit{3}}^{{++}}\quad{\textit{4}}^{{++}}$                  \\
 \vdots    & \vdots       & \vdots        & \vdots         \\
                                                        \hline\hline
\end{tabular}
\caption{Quantum number ($J^{PC}$) of $J/\psi J/\psi$ system. The number in the bracket is in the $|JMLS\rangle$ representation, with the form of $^{2S+1}L_J$. \label{tab:JPC}}
\end{table}
%%%%--------------------
The quantum numbers in bold and italics in Table \ref{tab:JPC} are those that meet the requirements of discrete and Bose symmetry. Since the higher partial waves of $L\geq 2$ are ignored, the partial wave of italics is excluded, and there is no coupling between the partial waves of $L=0$ and $L=2$ (for example, $^1S_0-^5D_0$.).
%%%%
For the partial wave amplitudes $T^{ij}$ ($i\neq j$, inelastic scatterings), there are 41 independent helicity amplitudes, and the partial waves can be expressed as

\begin{eqnarray}\label{eq:pw41}
  T_{^1S_0}^{ij}(s)&=&\frac{1}{3}[2T_{++++}^{0,ij}(s)+2T_{++--}^{0,ij}(s)-2T_{++00}^{0,ij}(s)-2T_{00++}^{0,ij}(s)+T_{0000}^{0,ij}(s)]\,, \nonumber\\[3mm]
  %%%%%%
  T_{^5S_2}^{ij}(s)&=&\frac{1}{15}[T_{++++}^{2,ij}(s)+T_{++--}^{2,ij}(s)]+\frac{\sqrt{6}}{15}[T_{+++-}^{2,ij}(s)+T_{+-++}^{2,ij}(s)+T_{++-+}^{2,ij}(s)+T_{-+++}^{2,ij}(s)]\nonumber\\
  &&+\frac{\sqrt{3}}{15}[T_{+++0}^{2,ij}(s)+T_{+0++}^{2,ij}(s)+T_{++0+}^{2,ij}(s)+T_{0+++}^{2,ij}(s)+T_{++-0}^{2,ij}(s)+T_{-0++}^{2,ij}(s)+T_{++0-}^{2,ij}(s)+T_{0-++}^{2,ij}(s)]\nonumber\\
  &&+\frac{1}{5}[T_{+00+}^{2,ij}(s)+T_{0++0}^{2,ij}(s)+T_{+00-}^{2,ij}(s)+T_{0-+0}^{2,ij}(s)+T_{+0+0}^{2,ij}(s)+T_{0+0+}^{2,ij}(s)+T_{0+0-}^{2,ij}(s)+T_{+0-0}^{2,ij}(s)]\nonumber\\
  &&+\frac{\sqrt{2}}{5}[T_{+-+0}^{2,ij}(s)+T_{+0+-}^{2,ij}(s)+T_{+-0+}^{2,ij}(s)+T_{0++-}^{2,ij}(s)+T_{-++0}^{2,ij}(s)+T_{+0-+}^{2,ij}(s)+T_{-+0+}^{2,ij}(s)+T_{0+-+}^{2,ij}(s)]\nonumber\\
  &&+\frac{2}{15}[T_{++00}^{2,ij}(s)+T_{00++}^{2,ij}(s)+T_{0000}^{2,ij}(s)]+\frac{2\sqrt{6}}{15}[T_{+-00}^{2,ij}(s)+T_{00+-}^{2,ij}(s)]+\frac{2}{5}[T_{+-+-}^{2,ij}(s)+T_{+--+}^{2,ij}(s)]\nonumber\\
  &&+\frac{2\sqrt{3}}{15}[T_{+000}^{2,ij}(s)+T_{00+0}^{2,ij}(s)+T_{0+00}^{2,ij}(s)+T_{000+}^{2,ij}(s)]\,, \nonumber\\[3mm]
  %%%%%%
  T_{^3P_0}^{ij}(s)&=&T_{++++}^{0,ij}(s)-T_{++--}^{0,ij}(s)\,, \nonumber\\[3mm]
  %%%%%%
  T_{^3P_1}^{ij}(s)&=&\frac{1}{2}[T_{+0+0}^{1,ij}(s)-T_{+00+}^{1,ij}(s)-T_{0++0}^{1,ij}(s)+T_{+0-0}^{1,ij}(s)-T_{+00-}^{1,ij}(s)-T_{0-+0}^{1,ij}(s)+T_{0+0+}^{1,ij}(s)+T_{0+0-}^{1,ij}(s)]\,, \nonumber\\[3mm]
  %%%%%%
  T_{^3P_2}^{ij}(s)&=&\frac{\sqrt{3}}{5}[T_{+++0}^{2,ij}(s)+T_{+0++}^{2,ij}(s)+T_{++0+}^{2,ij}(s)+T_{0+++}^{2,ij}(s)-T_{++-0}^{2,ij}(s)-T_{-0++}^{2,ij}(s)-T_{++0-}^{2,ij}(s)-T_{0-++}^{2,ij}(s)]\nonumber\\
  &&+\frac{3}{10}[T_{+0+0}^{2,ij}(s)+T_{0+0+}^{2,ij}(s)-T_{0+0-}^{2,ij}(s)-T_{-0+0}^{2,ij}(s)+T_{+00+}^{2,ij}(s)+T_{0++0}^{2,ij}(s)-T_{+00-}^{2,ij}(s)-T_{0-+0}^{2,ij}(s)]\nonumber\\
  &&+\frac{2}{5}[T_{++++}^{2,ij}(s)-T_{++--}^{2,ij}(s)]\,. 
\end{eqnarray}
Due to the time reversal invariance, the partial wave amplitudes of $T^{ii}$ (elastic scatterings), i.e., the amplitudes of processes of $J/\psi J/\psi\to J/\psi J/\psi, J/\psi \psi(2S)\to J/\psi \psi(2S), J/\psi \psi(3770)\to J/\psi \psi(3770)$, have only 25 independent helicity amplitudes and the partial waves can be simplified as
\begin{eqnarray}\label{eq:pw25}
  T_{^1S_0}^{ii}(s)&=&\frac{1}{3}[2T_{++++}^{0,ii}(s)+2T_{++--}^{0,ii}(s)-4T_{++00}^{0,ii}(s)+T_{0000}^{0,ii}(s)]\,, \nonumber\\[3mm]
  %%%%%%
  T_{^5S_2}^{ii}(s)&=&\frac{2\sqrt{6}}{15}[T_{+++-}^{2,ii}(s)+T_{++-+}^{2,ii}(s)]+\frac{2\sqrt{3}}{15}[T_{+++0}^{2,ii}(s)+T_{++0+}^{2,ii}(s)+T_{++-0}^{2,ii}(s)+T_{++0-}^{2,ii}(s)]+\frac{2}{15}T_{0000}^{2,ii}(s)\nonumber\\
  &&+\frac{2}{5}[T_{+-+-}^{2,ii}(s)+T_{+--+}^{2,ii}(s)+T_{+00+}^{2,ii}(s)+T_{+00-}^{2,ii}(s)]+\frac{2\sqrt{2}}{5}[T_{+-+0}^{2,ii}(s)+T_{+-0+}^{2,ii}(s)+T_{-++0}^{2,ii}(s)+T_{-+0+}^{2,ii}(s)]\nonumber\\
  &&+\frac{4}{15}T_{++00}^{2,ii}(s)+\frac{1}{5}[T_{+0+0}^{2,ii}(s)+T_{0+0+}^{2,ii}(s)+T_{0+0-}^{2,ii}(s)+T_{+0-0}^{2,ii}(s)]+\frac{4\sqrt{6}}{15}T_{+-00}^{2,ii}(s)\nonumber\\
  &&+\frac{4\sqrt{3}}{15}[T_{+000}^{2,ii}(s)+T_{0+00}^{2,ii}(s)]+\frac{1}{15}[T_{++++}^{2,ii}(s)+T_{++--}^{2,ii}(s)]\,, \nonumber\\[3mm]
  %%%%%%
  T_{^3P_0}^{ii}(s)&=&T_{++++}^{0,ii}(s)-T_{++--}^{0,ii}(s)\,, \nonumber\\[3mm]
  %%%%%%
  T_{^3P_1}^{ii}(s)&=&\frac{1}{2}[T_{+0+0}^{1,ii}(s)-2T_{+00+}^{1,ii}(s)+T_{+0-0}^{1,ii}(s)-2T_{+00-}^{1,ii}(s)+T_{0+0+}^{1,ii}(s)+T_{0+0-}^{1,ii}(s)]\,, \nonumber\\[3mm]
  %%%%%%
  T_{^3P_2}^{ii}(s)&=&\frac{2}{5}[T_{++++}^{2,ii}(s)-T_{++--}^{2,ii}(s)]+\frac{2\sqrt{3}}{5}[T_{+++0}^{2,ii}(s)+T_{++0+}^{2,ii}(s)-T_{++-0}^{2,ii}(s)-T_{++0-}^{2,ii}(s)]\nonumber\\
  &&+\frac{3}{10}[T_{+0+0}^{2,ii}(s)+T_{0+0+}^{2,ii}(s)-T_{0+0-}^{2,ii}(s)-T_{-0+0}^{2,ii}(s)]+\frac{3}{5}[T_{+00+}^{2,ii}(s)-T_{+00-}^{2,ii}(s)]\,. 
\end{eqnarray}
For details of each partial wave, see Eq.(\ref{eq:pw amp}) in Appendix~\ref{app:1}.

\subsection{Unitarizaton}
The unitarity  of the partial wave amplitudes in terms of the $|JMLS\rangle$ representation is given as  \cite{Zhou:2022xpd,Martin1970,Oller:2019opk,Chung:1971ri}.
$|\vec{p}~^{\prime\prime}|, E''_{cm}$ are the modulus of the three-momentum and energy of one of the particles in the intermediate process.
%%%%
As can be found in Table \ref{tab:JPC}, there is no coupled partial waves with different initial and final orbit angular momentums, where $L$ should be either zero or one. 
%(such as $^1S0-^5D_0$) 
In the present analysis, we consider the coupled channels scattering of $J/\psi J/\psi$-$J/\psi \psi(2S)$-$J/\psi \psi(3770)$. Therefore, one can write the unitarity relation for each partial wave as 
\begin{eqnarray}
  {\rm Im} T^{ij}_{JLS}&=&\sum_{k=1}^a T^{ik}_{JLS}~\rho_k ~T^{kj~*}_{JLS}\,, \label{eq:unit;3}
 \end{eqnarray}
 \begin{eqnarray}
\rho_k(s)=\frac{2|\vec{p}_k|}{E_k}=\frac{\lambda^{1/2}(s,m_{1k}^2, m_{2k}^2)}{s} \,.\label{eq;rho}
\end{eqnarray}
where \lq a=2,3' represents a couple-channels case or a triple-channels case. See discussions below. For simplicity, the quantum numbers \lq J, L, S' are ignored from now on. $\rho_k$ is the phase space factor for the $k$-th channel, given as \cite{Kuang:2020bnk}, where $m_{1k}$ and $m_{2k}$ are the masses of the two particles in the $k$-th channel. 
The triangle function $\lambda(s,m_{1k}^2, m_{2k}^2)$ has been given in the Appendix.\ref{app:1}.
%%%%%

The scattering amplitudes given in Eqs.~(\ref{eq:pw41},\ref{eq:pw25}) do not fulfill unitarity since they are calculated according to the spirit of perturbation theory. 
In order to restore the unitary,  we apply Pad\'e  approximation 
\cite{Truong:1988zp,Dai:2011bs,Dai:2012kf} to realize the unitarization\footnote{There are some similar approaches are successful in unitarizing chiral amplitudes, for instance, the inverse amplitude method. See Refs.~\cite{Dobado:1996ps,Oller:1997ng,GomezNicola:2001as} for details.}. Here, the matrix Pad\'e approximation is performed by constructing the amplitudes from the LO and NLO amplitudes. One has
\begin{eqnarray}
   T=T^{LO}\cdot[T^{LO}-T^{NLO}]^{-1}\cdot T^{LO} \,. \label{eq:pade}
\end{eqnarray}
It is not hard to check that Eq.~\eqref{eq:pade} can not only satisfy coupled channel unitarity as given in Eq.~\eqref{eq:unit;3}, but also restore the perturbation calculations up to NLO once the one-loop corrections are smaller than that of the tree diagrams\footnote{We are aware that the Pad\'e approximation is not as model-independent as some other methods such as dispersion relation \cite{Dai:2014zta,Dai:2016ytz}, and it violates the crossing symmetry and introduce some fake poles \cite{Dai:2011bs,Dai:2012kf}. However, it is still successful in confirming the existence of the $\sigma$ and $\kappa$, with reasonable poles found by unitarizing amplitudes of chiral perturbation theory.  }. 
%%%%%
In practice, the partial waves of $^3P_0$ and $^3P_2$ (with the quantum numbers $0^{-+}$ and $2^{-+}$) vanish at LO, and they are still small at NLO. Hence, we do not perform unitarization on them to avoid further complications to our model. Finally, we apply the Pad\'e approximation for three partial waves, $^1S_0$, $^5S_2$, $^3P_1$. Furthermore, for $^3P_0$ and $^3P_2$ waves, we use the perturbative amplitudes without unitarization. 
We sum these five partial waves to obtain the invariant mass spectrum of $J/\psi$-pair system and fit it to the data. 
%%%
The relation between $S$ matrix element and $T$ amplitudes may also be helpful for searching poles: 
\begin{eqnarray}
   S_{jk}(s)&=&\delta^j_k+2i\sqrt{\rho_j(s)\rho_k(s)}T_{jk}(s) \,.  \label{eq:S&T}
\end{eqnarray}
With the fixed $T$ amplitudes and $S$ matrix elements, one can extract the pole information and study the property of the resonance. See discussions in the next section.

\section{Fit results and discussions}\label{Sec:III}
\subsection{Fit to the invariant mass spectra}
With the partial wave amplitudes obtained by Pad\'e approximation, one can get the helicity amplitude and fit it to the $J/\psi J/\psi$ invariant mass spectrum. The events distribution is calculated by \cite{Dai:2021wxi}
\begin{eqnarray}
   \frac{d~{\rm Events^1}}{d\sqrt{s}}&=&\tilde{N}_1 ~p_{cm}(s) \sum_{ \mu_1\mu_2 \mu_3\mu_4}   \int_{-1}^{1}dz_s  |\sum_{i=1}^{a}\alpha_i T^{i1}_{ \mu_1\mu_2 \mu_3\mu_4}  (s,z_s)|^2 \,, \nonumber\\
   \frac{d~{\rm Events^2}}{d\sqrt{s}}&=&\tilde{N}_2 ~p_{cm}^\prime(s) \sum_{ \mu_1\mu_2 \mu_3\mu_4}   \int_{-1}^{1}dz_s  |\sum_{i=1}^{a}\alpha_i T^{i2}_{ \mu_1\mu_2 \mu_3\mu_4}  (s,z_s)|^2 \,,        \label{eq:events}
\end{eqnarray} 
where $p_{cm}$ and $p_{cm}^\prime$ are the momentum of $J/\psi J/\psi$ and $J/\psi \psi(2S)$ in the center-of-mass frame, respectively. The superscript of the summation symbol is the number of coupled channels we consider, i.e., a = 2 for a couple-channels case and a = 3 for a triple-channels case. The superscripts \lq 1,2' represent for invariant mass spectra of di-$J/\psi$ or $J/\psi \psi(2S)$.  
Note that only ATLAS gives the data of the $J/\psi \psi(2S)$ invariant mass spectra, which will be able to perform a combined analysis on these two invariant mass spectra.
$\tilde{N}_{1,2}$ is a normalization factor. Notice that the other factors, such as the integration on the azimuthal angle $\phi$, a factor $2\pi$, and the normalization factor of the final states, have been absorbed into the normalization factor $\tilde{N}_{1,2}$. The superscript \lq i' is the label for the channels. See discussions below.  
As discussed before, the helicity amplitudes are composed of five partial waves, $F_{^1S_0}(s)$, $F_{^5S_2}(s)$, $F_{^3P_0}(s)$, $F_{^3P_1}(s)$, $F_{^3P_2}(s)$. 
The quantitative contributions of the intermediate states $J/\psi J/\psi$ to  $J/\psi J/\psi$ re-scattering are unknown. Indeed it is possible that each of the amplitudes, $T^{11}$, $T^{21}$, and $T^{31}$, has a significant contribution to the $J/\psi J/\psi$ invariant mass spectrum. In addition, the threshold of $J/\psi J/\psi(2S)$ is 6783 MeV, and the threshold of $J/\psi J/\psi(3770)$ is 6867.6 MeV, which is close to the $X(6900)$ and may play a significant role on the resonance structure in the invariant mass spectrum.    
Hence, the strategy we use here is that: each channel, $T_{i1}$ ($T_{i2}$), contributes a ratio, $\alpha_i$, with the normalization condition $\sum_i\alpha_i^2\equiv 1$. Note that the $\alpha_i^2$ and $\tilde{N}$ are multiplied together and fitted to the distribution of the events.
Hence the normalization condition is to fix the dependence between them. 
The specific expression of the amplitude part of Eq.\eqref{eq:events} can then be expressed as 
\begin{eqnarray}
  &&\sum_{\footnotesize \mu_1\mu_2 \mu_3\mu_4}     \int_{-1}^{1}|\sum_{i=1}^{a}\alpha_i T^{i1,i2}_{\footnotesize \mu_1\mu_2 \mu_3\mu_4}  (s,z_s)|^2 dz_s 
  =512\pi^2
  \left[~|F^{1,2}_{^1S_0}(s)|^2 
   +5|F^{1,2}_{^5S_2}(s)|^2+|F^{1,2}_{^3P_0}(s)|^2 +3|F^{1,2}_{^3P_1}(s)|^2+5|F^{1,2}_{^3P_2}(s)|^2~
  \right] \, , \nonumber\\
  \label{eq:mod;T}
\end{eqnarray}
The relationship between F amplitudes and T amplitudes can be expressed as
% \begin{eqnarray}\label{eq:FJLS}
%     F^{1,2}_{JLS}(s)=\sum_{i=1}^a\alpha_i N_i T^{i1,i2}_{JLS}(s)
% \end{eqnarray}
\begin{eqnarray}\label{eq:FJLS}
    F^{1}_{JLS}(s)&=&\alpha_1 N_1 T^{11}_{JLS}(s) + \alpha_2 N_2 T^{21}_{JLS}(s) + \alpha_3 N_3 T^{31}_{JLS}(s)\,, \nonumber\\
    F^{2}_{JLS}(s)&=&\alpha_1 N_1 T^{12}_{JLS}(s) + \alpha_2 N_2 T^{22}_{JLS}(s) + \alpha_3 N_3 T^{32}_{JLS}(s)\,,    
\end{eqnarray}
Here $N_{i}$ is the normalization factor caused by the property of identical particles in the initial states of $T$ amplitudes, which is given as 
$N_1=\sqrt{2}$ and $N_{2,3}=1$.  
Indeed, the way to achieve $F^{1,2}_{JLS}(s)$ is indeed the same as the Au-Morgan-Pennington (AMP) method \cite{Au:1986vs,Dai:2014lza}, where final state interactions \cite{Yao:2020bxx} is taken into account systematically, with $\alpha_i$ including the left-hand cut and distant right-hand cut.

As discussed above, except for measurements on the invariant mass spectrum of $J/\psi$ pairs by LHCb collaboration \cite{LHCb:2020bwg}, there are two other new measurements. One is from the CMS collaboration \cite{Zhang:2022toq}, where the X(6900) is confirmed in the invariant mass of di-$J/\psi$, and also a new X(6600) resonant structure is observed. The other is from ATLAS collaboration \cite{Xu:2022rnl},  they measured the invariant mass spectra of 4$\mu$, both from di-$J/\psi$ and from $J/\psi$-$\psi(2S)$, respectively.  
%%% 
Following these three sets of data, we perform three kinds of fits. Each focuses on one data set, and each has two fits: one for couple-channels case ($J/\psi J/\psi$-$J/\psi \psi(2S)$) and the other for triple-channels case ($J/\psi J/\psi$-$J/\psi \psi(2S)$-$J/\psi \psi(3770)$).  \footnote{In Ref.\cite{Zhou:2022xpd}, we have already presented the results with LHCb's experimental data. Here we add the fits for the data sets of CMS and ATLAS to check the stability of the conclusion. } At the end of the day, we classify our fits as follows:  the ones for LHCb (Fit.~I for the couple-channels case and Fit.~IV for the triple-channels case), for CMS (Fit.~II and Fit.~V),  and for ATLAS (Fit.~III and Fit.~VI).

\begin{figure}[htbp]
   \centering
   {\includegraphics[width=0.9\linewidth, height=0.8\textheight]{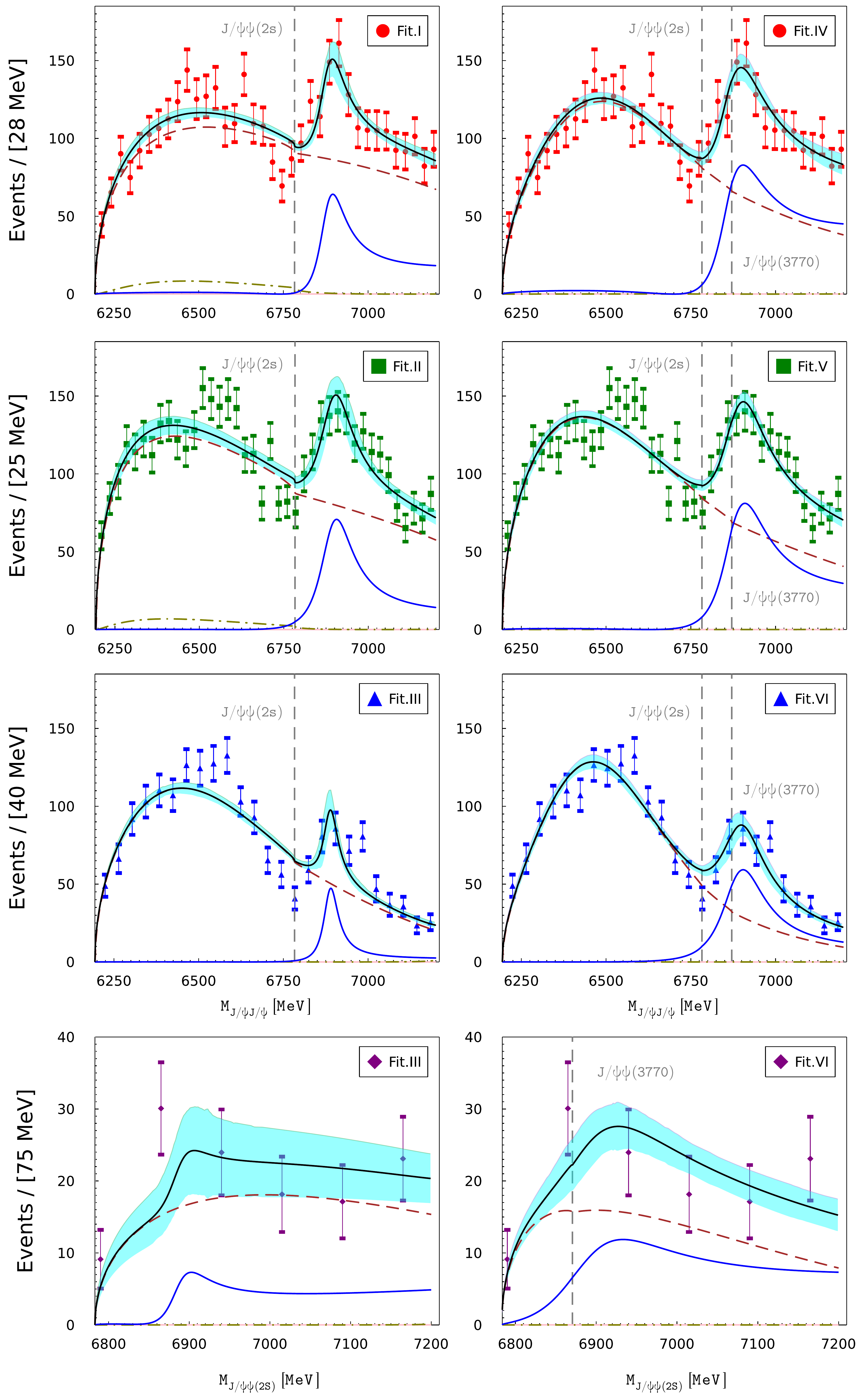}} 
   \caption{Fits to the invariant mass spectra and individual contribution of each partial wave. The graphs on the left side are for the couple-channels case, and the ones on the right side are for the triple-channels case.
   The data sets are taken from Refs. \cite{LHCb:2020bwg,Zhang:2022toq,Xu:2022rnl}.  The cyan bands are the uncertainties of our solutions, taken from the bootstrap method within 2$\sigma$. \label{Fig:IMS}}
\end{figure}

The input parameters, such as the masses of the particles, are taken from PDG \cite{ParticleDataGroup:2020ssz}. They are given as: $m_{J/\psi}=3096.9$~MeV, $m_{\psi(2S)}=3686.1$~MeV, $m_{\psi(3770)}=3773.7$~MeV. 
The renormalization scale of one-loop amplitudes is taken to be $\mu=1$~GeV. 
The other parameters, the couplings of the effective Lagrangians and the normalization factor, are
fixed by MINUIT \cite{James:1975dr}, which is a common tool to find the solution with minimum $\chi^2_{d.o.f.}$. The parameters and the $\chi^2_{\rm d.o.f}$ of Fits. I, II, and III are shown in Table \ref{tab:para two} for the couple-channels case. 
%%%-----------------------------------------------
\begin{table}[htbp]
   %   \caption{Fitted parameters.  }
   {\footnotesize  
      \begin{tabular}{cccc}
      \hline\hline
      parameter  & $\mathbf{Fit.I(LHCb)}$  &  Fit.II(CMS)    & Fit.III(ATLAS)                     \\ \hline 
      $c_1$      & $-0.1236_{-0.0001}^{+0.0001}$     &$-0.1504^{+0.0001}_{-0.0002}$      & $-0.0618_{-0.0001}^{+0.0001}$      \\      
      $c_2$      & $-0.5336_{-0.0001}^{+0.0021}$   &$-0.6203^{+0.0001}_{-0.0001}$ & $-0.3369_{-0.0001}^{+0.0004} $     \\
      $c_3$      & $-0.3180_{-0.0001}^{+0.0171}$  &$-0.3492^{+0.0004}_{-0.0001}$ & $-0.3171_{-0.0003}^{+0.0091} $     \\
      $c_4$      & $-0.6178_{-0.0002}^{+0.0234}$  &$-0.6835^{+0.0004}_{-0.0001}$  &$-0.5386_{-0.0001}^{+0.0078}   $      \\
      $\tilde{N_1}$ & ~~$1.5600_{-0.0850}^{+0.6284}$   &$~~0.5336^{+0.1404}_{-0.0193}$ &~~$0.1888_{-0.1934}^{+0.1109}$    \\
    $\tilde{N_2}$ & $\cdots$                       &$\cdots$                           &~~$0.2200_{-0.0198}^{+0.0721}$    \\
      $\alpha_1$ & ~~$0.3831_{-0.0052}^{+0.0104}$  &$~~0.3510^{+0.0012}_{-0.0001}$     & ~~$0.1812_{-0.0032}^{+0.0473}$    \\
     $\alpha_2$ & $-0.9237_{-0.0022}^{+0.0089}$  &$-0.9364^{+0.0001}_{-0.0009}$       & $~~0.9834_{-0.0088}^{+0.1861}$         \\
      $\chi^2_{d.o.f.}$    &   1.31                  &$1.77$                            &2.53                               \\
      \hline\hline
      \end{tabular}
      \caption{Parameters of couple-channels case. The unit of the normalization factor $\tilde{N}$ is $10^{-4}{\rm MeV}^{-2}$. The uncertainties of the parameters are taken from bootstrap.  \label{tab:para two}  }
   }       
      \end{table}
%%%-----------------------------------------------
%%%%%%%%%%%%%%%%%
Those of Fits.~IV, V, and VI are shown in  Table\ref{tab:para three} for the triple-channels case. 
%%%-----------------------------------------------
\begin{table}[htbp]
   %   \caption{Fitted parameters.  }
{\footnotesize  
   \begin{tabular}{ccccc}
   \hline\hline
   parameter   &  $\mathbf{Fit.IV(LHCb)}$    & Fit.V(CMS)  & Fit.VI(ATLAS)                     \\ \hline 
   $c_1$       & $-0.1254_{-0.0002}^{+0.0007}$                 &$-0.1466^{+0.0007}_{-0.0001}$     &$-0.1258^{+0.0001}_{-0.0001}$    \\[0.75mm]   
   $c_2$        & $-0.5860_{-0.0001}^{+0.0001} $                &$-0.5892^{+0.0002}_{-0.0001}$    &$-0.5895^{+0.0001}_{-0.0001}$    \\[0.75mm]
   $c_3$       & ~~$0.1908_{-0.0013}^{+0.0024} $               &$~~0.0072^{+0.0045}_{-0.0001  }$   &$~~0.0993^{+0.0008}_{-0.0001}$   \\[0.75mm]
   $c_4$        &$-1.0690_{-0.0022}^{+0.0055}   $              &$-0.8476^{+0.0069}_{-0.0001}$      &$-1.0289^{+0.0014}_{-0.0011}$    \\[0.75mm]
      $c_5$        & $ -0.0611_{-0.0001}^{+0.0001}$             &$-0.0892^{+0.0006}_{-0.0001}$     &$-0.0542^{+0.0001}_{-0.0001}$    \\[0.7mm]
      $c_6$          & $-0.2811_{-0.0003}^{+0.0006}  $           &$-0.3468^{+0.0016}_{-0.0010}$     &$-0.2897^{+0.0002}_{-0.0001}$    \\[0.75mm]
      $c_7$          & ~~$0.5994_{-0.0003}^{+0.0007} $           &$~~0.7950^{+0.0020}_{-0.0017}$     &$~~0.5903^{+0.0002}_{-0.0001}$   \\[0.75mm]
      $c_8$         &~~$0.2618_{-0.0001}^{+0.0003}  $            &$~~0.5295^{+0.0004}_{-0.0001}$     &$~~0.2299^{+0.0001}_{-0.0001}$   \\[0.75mm]
      $c_9$          &$-0.2169_{-0.0001}^{+0.0007}  $             &$-0.6789^{+0.0027}_{-0.0001}$     &$-0.2408^{+0.0002}_{-0.0001}$    \\[0.75mm]
      $\tilde{N_{1}}$    &~~$2.4583_{-0.2999}^{+1.5452}$           &$~~0.7274^{+0.2454}_{-0.1157}$     &~~$0.1487^{+0.0750}_{-0.0214}$    \\[0.7mm]
      $\tilde{N_{2}}$    &$\cdots$                                  &$\cdots$                        &~~$0.1247^{+0.0651}_{-0.0184}$    \\[0.7mm]
      $\alpha_1$   & ~~$0.3624_{-0.0254}^{+0.0529}$               &$~~0.2721^{+0.0018}_{-0.0003}$      &$~~0.1761^{+0.0252}_{-0.0083}$    \\[0.75mm]    
     $\alpha_2$   & $-0.8610_{-0.0511}^{+0.1072}$                &$-0.9312^{+0.0276}_{-0.0088}$        &$-0.7333^{+0.1380}_{-0.0426}$                \\[0.75mm]
      $\alpha_3$   & $-0.3568_{-0.0633}^{+0.1085} $                 &$-0.2426^{+0.0912}_{-0.0511}$       &$-0.6567^{+0.2220}_{-0.0901}$    \\[0.75mm]
      $\chi^2_{d.o.f.}$         &1.30                                   &$1.95$                          &$1.91$                          \\
   \hline\hline
   \end{tabular}
   \caption{Parameters of triple-channels case. The unit of the normalization factor $\tilde{N}$ is $10^{-4}{\rm MeV}^{-2}$. The uncertainties of the parameters are taken from bootstrap.  \label{tab:para three}  }
   }       
\end{table}
%%%-----------------------------------------------
The errors of the parameters are mainly from bootstrap \cite{Efron:1979bxm} rather than MINUIT, as the latter is much smaller. The uncertainties of bootstrap are counted by varying the experimental data within its uncertainty, multiplying a normal distribution function. The fit results are shown in Fig.\ref{Fig:IMS}. 
%%%---------------------------------------------------
%%%%%%%%%%%%%%------------------------------------------

It can be found from Tables  \ref{tab:para two} and \ref{tab:para three} that, almost in all the Fits, $c_2 $, $c_4 $ and $c_7 $ are relatively large, indicating that $T^{12}$, $T^{22}$, and $T^{33}$ play significant roles in the coupled channel scatterings. They could contribute to the di-$J/\psi$ spectrum through process such as $J/\psi \psi(2S)\to J/\psi \psi(2S) \to J/\psi J/\psi$.
This implies that $J/\psi \psi(2S)$ should have a relatively significant contribution to the resonant structure of the $X(6900)$. The strength of the couplings confirms this point. See discussions on the residues of the $X(6900)$.

As can be seen in Fig.\ref{Fig:IMS}, the solutions of the triple-channels case fit better to the data in the energy region from 6200 MeV to 6800 MeV than those of the couple-channels Fits. In comparison, both of them fit perfectly around 6900 MeV (the resonant structure for the $X(6900)$) except for Fits.III and VI, the couple-channels cases for ATLAS. Of course, The $\chi^2_{d.o.f.}$ (except for that of the ATLAS's) of couple-channels and triple-channels are almost the same, though the latter has better fit quality.  
This is caused by the fact that there are more parameters in the triple-channels case. Correspondingly, where the contributions from re-scatterings such as $J/\psi \psi(3770)\to J/\psi J/\psi$ and $J/\psi \psi(3770)\to J/\psi \psi(2S)$ have been included. It should be stressed that though the $J/\psi \psi(3770)$ channel will contribute little to the $X(6900)$, it supplies a significant background to the $J/\psi J/\psi$ invariant mass spectra and thus improve the fit quality, especially around the $J/\psi \psi(3770)$ threshold.    

%%%%%
Our Fits.~I and IV for LHCb data have much smaller $\chi^2_{d.o.f.}$. The reason is as follows: For the data of ATLAS, it has fewer statistics and data points. Not to say that some of the data points are not so consistent with each other in the energy region around the $X(6900)$;  For CMS's data, it has an apparent \lq peak-like' structure in the energy region around 6500~MeV, but our solutions do not have such a structure. Hence, the $\chi^2_{d.o.f.}$ is larger. Of course, our main goal is to extract the pole information and the quantum number of the resonance. Hence, we will satisfy the fit around 6500~MeV and pay more attention to the energy region around 6900~MeV\footnote{Indeed, our solutions do not find a pole like the $X(6600)$. It should be caused by either more complicated dynamics such as two-loop contributions or contributions from some other channels.}.  
%%%%
Roughly, the solutions of Fits.~I/IV (for LHCb) and Fit.II/V (for CMS) are similar to each other, as shown in the four graphs at the top of Fig.\ref{Fig:IMS}. This is confirmed by the relatively small differences between their parameters. See Tables~\ref{tab:para two} and \ref{tab:para three}.  For the solutions of ATLAS, they fit well the invariant mass spectra of di-$J/\psi$ and $J/\psi\psi(2S)$. See the last four graphs at the bottom of Fig.\ref{Fig:IMS}. 
Nevertheless, the latter has fewer statistics, and we will focus on the former to extract pole information. See discussions in the following sub-section.

%%%%

\subsection{On the nature of the \texorpdfstring{$X(6900)$}{} }
To study the nature of the $X(6900)$, one needs to extract the pole locations (mass and width) and quantum numbers from the partial wave scattering amplitudes. 
Firstly we need to extend the partial wave amplitudes to the complex-$s$ plane. It is performed by unitarity and reflection, and now searching for poles is converted into finding zeros of $S_{11}$, $S_{22}$, ${\rm det} S$, etc. See Ref.~\cite{Kuang:2020bnk} for details of the definition of nonphysical Riemann sheets. 
One also needs to search for poles in the relevant partial waves, i.e., $^1S_0$, $^5S_2$, and $^3P_1$ waves, while the other two ($^3P_0$ and $^3P_2$ waves) are rather small and contribute as a background. Finally, only one pole is found in the $^1S_0$ partial wave. All the pole parameters are shown in Tables \ref{tab:poles;2} and \ref{tab:poles;3} for couple-channels and triple-channels case, respectively.
%%%-----------------------------------------------------
\begin{table}[t]
{\footnotesize
\begin{tabular}{|c|c|c|cc|cc|}
\hline
\multirow{2}{*}{\rule[0cm]{0cm}{5mm}Data}&\multirow{2}{*}{\rule[0cm]{0cm}{5mm}RS} & \multirow{2}{*}{\rule[0cm]{0cm}{5mm}pole location (MeV)} & \multicolumn{2}{c|}{\rule[-0.2cm]{0cm}{5mm}$g_{J/\psi J/\psi}=|g|e^{i\varphi}$}  & \multicolumn{2}{c|}{$g_{J/\psi \psi(2S)}=|g|e^{i\varphi}$}    \\ \cline{4-7}  
&  &  &\multicolumn{1}{c|}{$|g_1|$(MeV)}  & \rule[-0.2cm]{0cm}{5mm}$\varphi_1(^\circ)$ & \multicolumn{1}{c|}{$|g_2|$(MeV)}  & $\varphi_2(^\circ)$  \\ \hline
\multirow{2}{*}{\rule[0cm]{0cm}{5mm}$\mathbf{LHCb(Fit.I)}$} &\rule[-0.2cm]{0cm}{7mm}II(- +)  & $6882.8_{-2.4}^{+8.3}$-$i16.5_{-0.3}^{+2.9} $  & \multicolumn{1}{c|}{$992.5_{-3.6}^{+36.5}$}  & $-89.9^{+0.8}_{-0.1}$  & \multicolumn{1}{c|}{$684.3_{-1.4}^{+4.0}$}  & $89.1^{+0.1}_{-0.1}$  \\ \cline{2-7}
&\rule[-0.2cm]{0cm}{7mm}III(- -)  & $6880.4_{-2.4}^{+8.4}$-$i27.8_{-0.1}^{+3.2}  $ & \multicolumn{1}{c|}{$981.2_{-3.8}^{+40.2}$} & $87.4^{+0.1}_{-0.1}$ & \multicolumn{1}{c|}{$681.0_{-2.2}^{+6.7}$} & $84.9^{+0.2}_{-0.1}$    
\\ \hline
\multirow{2}{*}{\rule[0cm]{0cm}{5mm}CMS(Fit.II)} &\rule[-0.2cm]{0cm}{7mm}II(- + )           & $6897.4_{-4.3}^{+10.7}$-$i20.9_{-0.3}^{+0.7} $  & \multicolumn{1}{c|}{$1113.7_{-4.2}^{+7.7}$}  & $89.9^{+0.1}_{-0.1}$                & \multicolumn{1}{c|}{$815.3_{-2.8}^{+4.7}$}  & $88.9^{+0.1}_{-0.1}$                             \\ \cline{2-7}
&\rule[-0.2cm]{0cm}{7mm}III(- -)          & $6893.2_{-4.1}^{+10.7}$ -$i37.7_{-0.1}^{+0.3}$   & \multicolumn{1}{c|}{$ 1104.2_{-4.8}^{+8.2}$} & $86.5^{+0.3}_{-0.1}$                & \multicolumn{1}{c|}{$804.5_{-2.8}^{+4.7}$}  & $83.8^{+0.1}_{-0.1}$                           \\ 
\hline
\multirow{2}{*}{\rule[0cm]{0cm}{5mm}ATLAS(Fit.III)} &\rule[-0.2cm]{0cm}{7mm}II(- + )           & $6887.3_{-2.9}^{+31.2}$-$i16.7_{-0.6}^{+0.2} $  & \multicolumn{1}{c|}{$821.1_{-13.3}^{+5.4}$}  & $-89.9^{+0.1}_{-0.1}$                & \multicolumn{1}{c|}{$394.5_{-6.9}^{+2.3}$}  & $88.7^{+0.1}_{-0.1}$                               \\ \cline{2-7}
&\rule[-0.2cm]{0cm}{7mm}III(- - )          & $6886.6_{-2.9}^{+31.3}$ -$i20.6_{-0.4}^{+0.2}$   & \multicolumn{1}{c|}{$ 820.8_{-4.8}^{+4.8}$} & $89.1^{+0.1}_{-0.1}$                & \multicolumn{1}{c|}{$393.0_{-2.4}^{+2.3}$}  & $85.8^{+0.1}_{-0.1}$                              \\
\hline
\end{tabular}
\caption{\label{tab:poles;2} Poles locations and residues for Fits.~I-III. RS represents the Riemann sheet, and the signs in the bracket are for the phase space factors, $\rho_1$, and $\rho_2$.} 
}
\end{table}
%%%------------------------------------------------------------
%%%%%%%%%%%%%%%%%%%%%%%
%%%------------------------------------------------------------
\begin{table}[t]
{\footnotesize
\begin{tabular}{|c|c|c|cc|cc|cc|}
\hline
\multirow{2}{*}{\rule[0cm]{0cm}{5mm}Data}&\multirow{2}{*}{\rule[0cm]{0cm}{5mm}RS} & \multirow{2}{*}{\rule[0cm]{0cm}{5mm}pole location (MeV)} & \multicolumn{2}{c|}{\rule[-0.2cm]{0cm}{5mm}$g_{J/\psi J/\psi}=|g|e^{i\varphi}$}  & \multicolumn{2}{c|}{$g_{J/\psi \psi(2S)}=|g|e^{i\varphi}$}  & \multicolumn{2}{c|}{$g_{J/\psi \psi(3770)}=|g|e^{i\varphi}$}  \\ \cline{4-9}  
&  &  &\multicolumn{1}{c|}{$|g_1|$(MeV)}  & \rule[-0.2cm]{0cm}{5mm}$\varphi_1(^\circ)$ & \multicolumn{1}{c|}{$|g_2|$(MeV)}  & $\varphi_2(^\circ)$ & \multicolumn{1}{c|}{$|g_3|$(MeV)}  & $\varphi_3(^\circ)$ \\ \hline
\multirow{4}{*}{\rule[0cm]{0cm}{12mm}$\mathbf{LHCb(Fit.IV)}$} 
&\rule[-0.2cm]{0cm}{7mm}II(- + +)           & $6874.8_{-5.8}^{+5.0}$-$i50.4_{-1.1}^{+1.7} $  & \multicolumn{1}{c|}{$1398.5_{-15.8}^{+21.6}$}  & $85.9^{+0.3}_{-0.1}$                & \multicolumn{1}{c|}{$962.1_{-10.9}^{+14.9}$}  & $84.6^{+0.1}_{-0.1}$                & \multicolumn{1}{c|}{$18.2_{-0.4}^{+0.7}$} & $-79.9^{+1.2}_{-0.2}$                \\ \cline{2-9}
& \rule[-0.2cm]{0cm}{7mm}III(- - +)          & $6862.0_{-6.2}^{+4.3}$ -$i68.9_{-2.0}^{+1.9}$ & \multicolumn{1}{c|}{$ 1364.7_{-12.2}^{+20.1}$} & $80.6^{+0.3}_{-0.1}$                & \multicolumn{1}{c|}{$927.4_{-8.3}^{+13.4}$} & $77.5^{+0.5}_{-0.1}$                & \multicolumn{1}{c|}{$19.3_{-0.4}^{+0.7}$} & $-79.0^{+0.6}_{-0.3}$  \\\cline{2-9}
& \rule[-0.2cm]{0cm}{7mm}IV(- - -)           & $6862.0_{-6.2}^{+4.3}$ -$i68.9_{-2.0}^{+1.9}$ & \multicolumn{1}{c|}{$ 1361.6_{-12.4}^{+19.0}$} & $80.7^{+0.2}_{-0.1}$                & \multicolumn{1}{c|}{$925.3_{-8.7}^{+12.5}$} & $77.5^{+0.4}_{-0.1}$                & \multicolumn{1}{c|}{$19.4_{-0.4}^{+0.7}$} & $-78.6^{+0.5}_{-0.2}$ \\ \cline{2-9}
&\rule[-0.2cm]{0cm}{7mm}VII(- + -)          & $6874.8_{-5.8}^{+5.0}$ -$i50.4_{-1.1}^{+1.7}$   & \multicolumn{1}{c|}{$ 1394.3_{-17.5}^{+17.7}$} & $85.9^{+0.2}_{-0.1}$                & \multicolumn{1}{c|}{$959.2_{-12.1}^{+11.7}$}  & $84.5^{+0.1}_{-0.1}$               & \multicolumn{1}{c|}{$18.4_{-0.4}^{+0.7}$} & $-79.2^{+1.0}_{-0.3}$               
\\ \hline
\multirow{4}{*}{\rule[0cm]{0cm}{12mm}CMS(Fit.V)} &\rule[-0.2cm]{0cm}{7mm}II(- + +)           & $6888.4_{-7.2}^{+11.3}$-$i59.4_{-0.5}^{+1.7} $  & \multicolumn{1}{c|}{$1452.8_{-6.8}^{+23.1}$}  & $85.6^{+0.1}_{-0.1}$                & \multicolumn{1}{c|}{$795.8_{-4.3}^{+12.2}$}  & $83.3^{+0.1}_{-0.1}$                & \multicolumn{1}{c|}{$38.8_{-0.1}^{+2.1}$} & $82.2^{+0.3}_{-0.1}$                \\ \cline{2-9}
&\rule[-0.2cm]{0cm}{7mm}III(- - +)          & $6878.9_{-7.4}^{+11.3}$ -$i73.1_{-1.1}^{+2.6}$ & \multicolumn{1}{c|}{$ 1430.3_{-5.7}^{+29.4}$} & $82.0^{+0.1}_{-0.1}$                & \multicolumn{1}{c|}{$773.9_{-4.2}^{+15.5}$} & $77.8^{+0.2}_{-0.1}$                & \multicolumn{1}{c|}{$36.4_{-0.1}^{+2.2}$} & $65.0^{+1.6}_{-0.4}$                \\ \cline{2-9}
&\rule[-0.2cm]{0cm}{7mm}IV(- - -)           & $6878.9_{-7.4}^{+11.3}$ -$i73.1_{-1.1}^{+2.6}$ & \multicolumn{1}{c|}{$ 1430.5_{-5.0}^{+18.8}$} & $82.0^{+0.1}_{-0.1}$                & \multicolumn{1}{c|}{$773.8_{-3.1}^{+8.7}$} & $77.8^{+0.2}_{-0.1}$                & \multicolumn{1}{c|}{$36.7_{-0.1}^{+2.1}$} & $65.6^{+1.6}_{-0.4}$                \\ \cline{2-9}
&\rule[-0.2cm]{0cm}{7mm}VII(- + -)          & $6888.4_{-7.2}^{+11.5}$ -$i59.4_{-0.5}^{+1.7}$   & \multicolumn{1}{c|}{$ 1452.3_{-5.6}^{+24.4}$} & $85.6^{+0.1}_{-0.1}$                & \multicolumn{1}{c|}{$795.4_{-3.4}^{+13.6}$}  & $83.3^{+0.1}_{-0.1}$               & \multicolumn{1}{c|}{$39.4_{-0.1}^{+2.2}$} & $83.4^{+0.4}_{-0.2}$                \\ \hline
\multirow{4}{*}{\rule[0cm]{0cm}{12mm}ATLAS(Fit.VI)} &\rule[-0.2cm]{0cm}{7mm}II(- + +)           & $6897.7_{-4.3}^{+19.1}$-$i50.9_{-0.2}^{+0.9} $  & \multicolumn{1}{c|}{$1409.8_{-1.9}^{+12.0}$}  & $86.2^{+0.1}_{-0.1}$                & \multicolumn{1}{c|}{$997.0_{-1.8}^{+8.8}$}  & $85.0^{+0.1}_{-0.1}$                & \multicolumn{1}{c|}{$5.7_{-0.1}^{+0.1}$} & $56.7^{+0.8}_{-0.3}$                \\ \cline{2-9}
&\rule[-0.2cm]{0cm}{7mm}III(- - +)          & $6883.8_{-4.0}^{+18.3}$ -$i73.4_{-0.7}^{+2.8}$ & \multicolumn{1}{c|}{$ 1373.6_{-2.7}^{+7.3}$} & $80.8^{+0.1}_{-0.1}$                & \multicolumn{1}{c|}{$960.0_{-1.3}^{+5.6}$} & $77.5^{+0.1}_{-0.2}$                & \multicolumn{1}{c|}{$7.2_{-0.1}^{+0.1}$} & $21.6^{+1.1}_{-1.0}$                \\ \cline{2-9}
&\rule[-0.2cm]{0cm}{7mm}IV(- - -)           & $6883.8_{-4.0}^{+18.3}$ -$i73.4_{-0.7}^{+2.8}$ & \multicolumn{1}{c|}{$ 1379.0_{-2.0}^{+10.0}$} & $80.8^{+0.1}_{-0.1}$                & \multicolumn{1}{c|}{$963.8_{-1.2}^{+7.1}$} & $77.5^{+0.1}_{-0.1}$                & \multicolumn{1}{c|}{$7.3_{-0.1}^{+0.1}$} & $22.1^{+1.1}_{-1.0}$                \\ \cline{2-9}
&\rule[-0.2cm]{0cm}{7mm}VII(- + -)          & $6897.7_{-4.3}^{+19.1}$ -$i50.9_{-0.2}^{+0.9}$   & \multicolumn{1}{c|}{$ 1406.7_{-2.0}^{+10.4}$} & $86.2^{+0.1}_{-0.1}$                & \multicolumn{1}{c|}{$994.9_{-2.3}^{+7.4}$}  & $85.0^{+0.1}_{-0.1}$               & \multicolumn{1}{c|}{$5.8_{-0.1}^{+0.2}$} & $57.6^{+0.9}_{-0.2}$                \\
\hline
\end{tabular}
\caption{\label{tab:poles;3} Poles locations and residues for Fits.~IV-VI. RS represents the Riemann sheets, and the signs in the bracket are for the phase space factors, $\rho_1$, $\rho_2$, and $\rho_3$. } 
}
\end{table}
%%%------------------------------------------------------------
Though the invariant mass spectra are quite different, it is impressive to find that the pole parameters are somehow stable in all these solutions. 
Firstly, we only find one resonant state in the $^1S_0$ (with quantum numbers of $0^{++}$) partial wave for each of these fits. Specifically, in each solution, we find two poles for the couple-channel case or four poles for the triple-channels case in the unphysical Riemann sheets (RS). Even the RSs (where the poles locate) are the same for these different solutions. This confirms the reliability of the models and also the extracted pole parameters. 

For the couple-channels case, two poles can be found in RS-II and RS-III in $^1S_0$ wave. The latter is the closest one to the physical sheet. The pole parameter of Fit.~I (fitting to the LHCb data) is $M=6880.4^{+8.4}_{-2.4}$ and $\Gamma = 55.6^{+6.4}_{-0.2}$, that of the Fit.~II (fitting to CMS's) is $M=6893.2^{+10.7}_{-4.1}$ and $\Gamma = 75.4^{+0.6}_{-0.2}$, and that of Fit.~III (fitting to ATLAS's)  is $M=6886.6^{+31.3}_{-2.9}$ and $\Gamma = 41.2^{+0.4}_{-0.8}$. According to the pole counting rule \cite{Morgan:1992ge,Dai:2011bs}, a pair of accompanying shadow poles in RS-II and RS-III indicate that the $X(6900)$ should be a Breit Wigner type particle. In another aspect, this resonance state contains at least four quarks ($cc\bar{c}\bar{c}$), so it is likely to be a compact tetra-quark state. Its couplings to the $J/\psi J/\psi$,  $J/\psi \psi(2S)$ channels are given in Table~\ref{tab:poles;2}. The magnitudes of $g_1$ and $g_2$ are large and in the same order. Both of them are much larger than that of $g_3$. It implies that the two channels, $J/\psi J/\psi$ and $J/\psi \psi(2S)$, couple strongly to the $X(6900)$. This is compatible with our discussions above, where $J/\psi \psi(2S)$ should have a relatively large contribution to the di-$J/\psi$ invariant mass spectrum.

For the triple-channels case, we find four poles in RS-II, RS-III, RS-IV, and RS-VII, and again in $^1S_0$ wave only. The information about their pole parameters is shown in Table \ref{tab:poles;3}.  
In the energy region between $\sqrt{s_{th_2}}=6783.0$~MeV and $\sqrt{s_{th_3}}=6870.6$~MeV,  RS-IV is the closest one to the physical sheet, while in the energy region above $\sqrt{s_{th_3}}=6867.6$~MeV,  RS-IV is the closest one to the physical sheet. 
Specifically, the poles being closet to the physical sheet are as follows: $M=6862.0^{+4.3}_{-6.2}$ and $\Gamma=137.8^{+3.8}_{-4.0}$ in RS-IV for Fit.~IV (fitting to LHCb's); $M=6878.9^{+11.3}_{-7.4}$ and $\Gamma=146.2^{+5.2}_{-2.2}$ in RS-IV for Fit.V (fitting to CMS's); And $M=6883.8^{+18.3}_{-4.0}$ and $\Gamma=146.8^{+5.6}_{-1.4}$ in RS-IV for Fit.VI (fitting to ATLAS's). 
It can also be seen from Table \ref{tab:poles;3} that for each fit, $g_3$ is much smaller than $g_1$ and $g_2$. This confirms that $J/\psi \psi(3770)$ should contribute little to the $X(6900)$. However, as we have pointed out in Ref.\cite{Zhou:2022xpd}, it still contributes to the amplitudes significantly as a background.  
This also demonstrates the correctness of our choice of a couple-channels model. 
Since there are four accompanying poles in the unphysical sheets, it again suggests that the $X(6900)$ should be a Breit- Wigner type particle. Further, it should be a compact tetra-quark state, similar to the conclusion of the couple-channels case.

We show the contribution of each partial wave in Fig.\ref{Fig:IMS}. 
The blue solid, brown dashed, gray dotted, olive dash-dotted, and pink solid lines are for $^1S_0$, $^5S_2$, $^3P_0$, $^3P_1$, and $^3P_2$ waves, respectively. As can be seen from Fig \ref{Fig:IMS}, for each solution, in the energy region between 6200~MeV and 6800~MeV, the contribution is mainly from the  $^5S_2$ partial wave, and $^1S_0$ and $^3P_1$ may have small contributions in some solutions. Also, no waves have a resonance-like structure around the 6600~MeV. This is compatible with the fact that we do not find a pole relative to the $X(6600)$.   
In contrast, in the energy region of [6800, 7200]~MeV, the main contribution is from $^1S_0$ and $^5S_2$, while the other contributions from the P-waves can be ignored. 
In all these solutions, the $^1S_0$ partial wave has an obvious resonant structure around 6900~MeV, and the $^5S_2$ partial wave contributes as a smooth background. It suggests that the $X(6900)$ should be $^1S_0(0^{++})$ state. The shape of the contribution of the  $X(6900)$ is similar to a normal Breit-Wigner's, being compatible with the conclusion from the pole counting rule. That is, the $X(6900)$ looks like a normal Breit-Wigner resonance.

\begin{figure}[htbp]
   \centering
   {\includegraphics[width=1\linewidth, height=0.4\textheight]{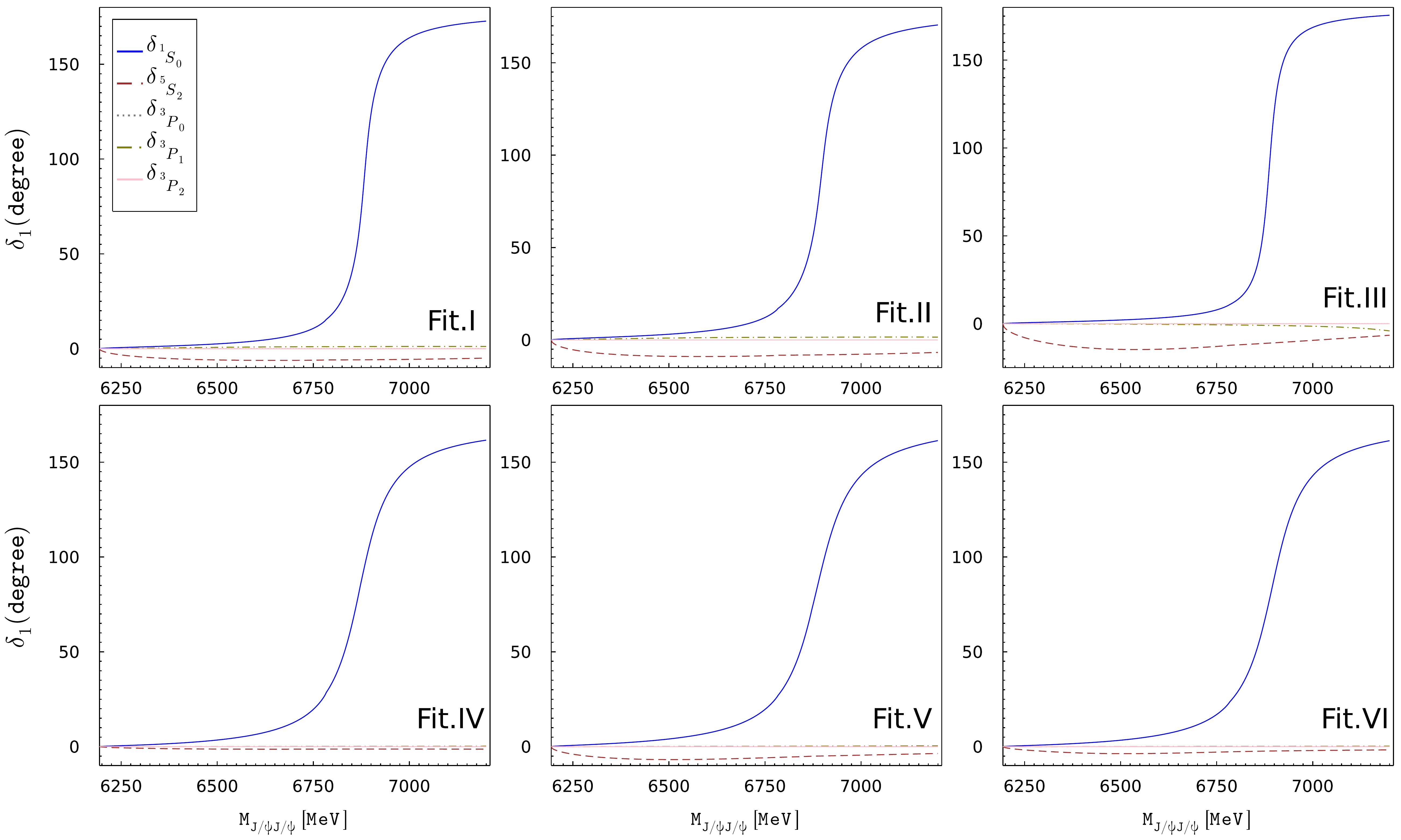}} 
   \caption{Phase shifts of $J/\psi J/\psi$ scatterings of different partial waves. The solid blue, dashed brown, dash-dotted olive, dotted gray, and solid pink lines are for $^1S_0$, $^5S_2$,  $^3P_1$, $^3P_0$, and $^3P_2$, respectively.   \label{fig:pahse-shift}  }
\end{figure}

As discussed in the introduction, the phase shifts help study the property of the state. With intuitive views, a narrow Breit-Wigner resonance should have a step-function-like phase shift of the scattering amplitude, which dramatically jumps from 0 to $\pi$.  
Therefore, we give the phase shifts of $\delta_1(J/\psi J/\psi\to J/\psi J/\psi)$ of each partial wave, shown in Fig.\ref{fig:pahse-shift}. 
%%%%%%%%%%%%%%
%%%%%%%%%%%%%%
For all the Fits, these phase shifts are similar to each other. The phase shift of the $^1S_0$ partial wave is very likely to be generated by a normal Breit-Wigner resonance, which rises 180 degrees steeply and crosses 90 degrees around 6900~MeV. The phase shifts of other partial waves are tiny, and they should contribute as smooth \lq backgrounds', which changes slowly. In contrast, we do not find any poles in these partial waves too. This again supports the hypothesis that the $X(6900)$ is a tetra-quark state with the quantum number of $0^{++}$.

\section{Summary}
\label{Sec:IV}
In this paper, we consider two models of coupled channel scatterings: $J/\psi J/\psi$ -$ J/\psi \psi(2S)$ and $J/\psi J/\psi$ -$ J/\psi \psi(2S)$-$ J/\psi \psi(3770)$. The effective Lagrangian is constructed, and the corresponding tree and one-loop Feynman diagrams for the scattering amplitudes are calculated. With these amplitudes, we perform partial wave decomposition to separate different partial wave amplitudes, where $^1S_0$, $^5S_2$, $^3P_0$,$^3P_1$, and $^3P_2$ waves are left. The pad\'e approximation is applied to recover the unitarity. By fitting to the $di$-$J/\psi$ invariant mass of LHCb, ATLAS, and CMS, as well as the $J/\psi \psi(2S)$ invariant mass spectrum from ATLAS, the unknown couplings are fixed. The amplitudes are continued to the complex-$s$ plane. From them, we extract the pole parameters for each partial wave. Our fits, coming from the triple-channels case, give the mass and width of the $X(6900)$ as: 
$M=6862.0^{+4.3}_{-6.2}$ and $\Gamma=137.8^{+3.8}_{-4.0}$ in Fit.IV, fitting to the date of LHCb; $M=6878.9^{+11.3}_{-7.4}$ and $\Gamma=146.2^{+5.2}_{-2.2}$ in Fit.V, fitting to the date of CMS; $M=6883.8^{+18.3}_{-4.0}$ and $\Gamma=146.8^{+5.6}_{-1.4}$ in Fit.VI, fitting to the date of ATLAS. All of these poles are from $^1S_0$ wave, implying that their quantum number is $0^{++}$. Since we find a pair of accompanying poles in the couple-channels model and four poles in the triple-channels model, the $X(6900)$ is likely to be a Breit-Wigner type particle, i.e., a compact tetra-quark.
We check it by extracting the phase shifts of each partial wave, and it supports the Breit-Wigner origin. Nevertheless, our solution can not describe the data around 6500~MeV well. Correspondingly, the $X(6600)$ is not found in our study, though it will not affect the results about the $X(6900)$.
To clarify the $X(6600)$, one needs more information about the dynamics in the relevant energy region. More experimental measurements on the angular distributions would lend credibility and plausibility to the partial wave decomposition and further study the nature of the $X(6900)$.

\section{Acknowledgements}
\label{Sec:V}
We are grateful to professors M. Shi and W. Shan for their helpful discussions. This work is supported by Joint Large Scale Scientific Facility Funds of the National Natural Science Foundation of China (NSFC) and Chinese Academy of Sciences (CAS) under Contract No.U1932110, NSFC Grants No.11805059, and Fundamental Research Funds for the central Universities.

\appendix
\section{Scattering amplitudes}\label{app:1}
The scattering amplitudes $T^{ij}$ can be expressed as
\begin{eqnarray}
    T^{ij}&=&F^{ij}_{(a)}(\varepsilon_1\cdot\varepsilon_2)(\varepsilon_3^*\cdot\varepsilon_4^*)+F^{ij}_{(b)}(\varepsilon_1\cdot\varepsilon_3^*)(\varepsilon_2\cdot\varepsilon_4^*)+F^{ij}_{(c)}(\varepsilon_1\cdot\varepsilon_4^*)(\varepsilon_2\cdot\varepsilon_3^*) \,, \label{eq:amp}
\end{eqnarray}
 The subscripts \lq 1,2,3,4' of the polarization vectors are the labels for the particles. The subscripts \lq (a,b,c)' are used to tag form factors with different polarization structures. Notice that the effective Lagrangians do not contain derivatives. Hence, the polarization vectors are contracted by the metric tensors $g^{\alpha\beta}g^{\alpha'\beta'}$ without complicated momentum terms, resulting in a simple formalism for the scattering amplitude as shown in Eq.~(\ref{eq:amp}). 
The specified expressions of the form factors $F^{ij}_{(k)}(k=a,b,c)$ up to NLO can be found in Eq.~\eqref{eq:amp loop coeff}.
The coupled channel scattering amplitudes up to the next-to-leading order are written as Eq.\eqref{eq:amp}. The form factors $F^{ij}_{(k)}$ are given by
 \begin{eqnarray}\label{eq:amp loop coeff}
   F^{11}_{(a)}&=&8c_1+\frac{1}{16\pi^2}[256c_1^2B_0(s,m_1^2,m_1^2)+32c_2^2B_0(s,m_1^2,m_2^2)+32c_5^2B_0(s,m_1^2,m_3^2)+(32c_3^2+16c_3c_4)B_0(s,m_2^2,m_2^2)\notag\\
   &&+(32c_6^2+16c_6c_7)B_0(s,m_3^2,m_3^2)+(16c_8^2+8c_8c_9)B_0(s,m_2^2,m_3^2)+64c_1^2B_0(t,m_1^2,m_1^2)+8c_2^2B_0(t,m_1^2,m_2^2)\notag\\
   &&+8c_5^2B_0(t,m_1^2,m_3^2)+4c_4^2B_0(t,m_2^2,m_2^2)+4c_7^2B_0(t,m_3^2,m_3^2)+2c_9^2B_0(t,m_2^2,m_3^2)+64c_1^2B_0(u,m_1^2,m_1^2)\notag\\
   &&+8c_2^2B_0(u,m_1^2,m_2^2)+8c_5^2B_0(u,m_1^2,m_3^2)+4c_4^2B_0(u,m_2^2,m_2^2)+4c_7^2B_0(u,m_3^2,m_3^2)+2c_9^2B_0(u,m_2^2,m_3^2)]\,, \nonumber\\
  %%%%
  F^{11}_{(b)}&=&8c_1+\frac{1}{16\pi^2}[8 c_2^2 B_0(s,m_1^2,m_2^2)+8 c_5^2 B_0(s,m_1^2,m_3^2)+64 c_1^2 B_0(s,m_1^2,m_1^2)+32 c_2^2 B_0(t,m_1^2,m_2^2)\notag\\
  &&+32 c_5^2 B_0(t,m_1^2,m_3^2)+256 c_1^2 B_0(t,m_1^2,m_1^2)+8 c_2^2 B_0(u,m_1^2,m_2^2)+8 c_5^2 B_0(u,m_1^2,m_3^2)+64 c_1^2 B_0(u,m_1^2,m_1^2)\notag\\
  &&+2 c_9^2 B_0(s,m_2^2,m_3^2)+4 c_4^2 B_0(s,m_2^2,m_2^2)+4 c_7^2 B_0(s,m_3^2,m_3^2)+8 c_8 (2 c_8+c_9) B_0(t,m_2^2,m_3^2)\notag\\
  &&+16 c_3 (2 c_3+c_4) B_0(t,m_2^2,m_2^2)+4 c_4^2 B_0(u,m_2^2,m_2^2)+16 c_6 (2 c_6+c_7) B_0(t,m_3^2,m_3^2)+2 c_9^2 B_0(u,m_2^2,m_3^2)\notag\\
  &&+4 c_7^2 B_0(u,m_3^2,m_3^2)]\,, \nonumber\\
  %%%%
  F^{11}_{(c)}&=&8c_1+\frac{1}{16\pi^2}[8 c_2^2 B_0(s,m_1^2,m_2^2)+8 c_5^2 B_0(s,m_1^2,m_3^2)+64 c_1^2 B_0(s,m_1^2,m_1^2)+8 c_2^2 B_0(t,m_1^2,m_2^2)+8 c_5^2 B_0(t,m_1^2,m_3^2)\notag\\
  &&+64 c_1^2 B_0(t,m_1^2,m_1^2)+32 c_2^2 B_0(u,m_1^2,m_2^2)+32 c_5^2 B_0(u,m_1^2,m_3^2)+256 c_1^2 B_0(u,m_1^2,m_1^2)+2 c_9^2 B_0(s,m_2^2,m_3^2)\notag\\
  &&+4 c_4^2 B_0(s,m_2^2,m_2^2)+4 c_7^2 B_0(s,m_3^2,m_3^2)+2 c_9^2 B_0(t,m_2^2,m_3^2)+4 c_4^2 B_0(t,m_2^2,m_2^2)+4 c_7^2 B_0(t,m_3^2,m_3^2)\notag\\
  &&+8 c_8 (2 c_8+c_9) B_0(u,m_2^2,m_3^2)+16 c_3 (2 c_3+c_4) B_0(u,m_2^2,m_2^2)+16 c_6 (2 c_6+c_7) B_0(u,m_3^2,m_3^2)]\,, \nonumber\\
  %%%%
  F^{12}_{(a)}&=&2c_2+\frac{1}{16\pi^2}[4 (2 c_3+7 c_4) c_2 B_0(s,m_1^2,m_2^2)+2 c_5 (2 c_8+7 c_9) B_0(s,m_1^2,m_3^2)+64 c_1 c_2 B_0(s,m_1^2,m_1^2)\nonumber\\
  &&+4 (2 c_3+c_4) c_2 B_0(t,m_1^2,m_2^2)+2 c_5 (2 c_8+c_9) B_0(t,m_1^2,m_3^2)+16 c_1 c_2 B_0(t,m_1^2,m_1^2)\nonumber\\
  &&+4 (2 c_3+c_4) c_2 B_0(u,m_1^2,m_2^2)+2 c_5 (2 c_8+c_9) B_0(u,m_1^2,m_3^2)
  +16 c_1 c_2 B_0(u,m_1^2,m_1^2)]\,, \nonumber\\
  %%%%
  F^{12}_{(b)}&=&2c_2+\frac{1}{16\pi^2}[4 (2 c_3+c_4) c_2 B_0(s,m_1^2,m_2^2)+2 c_5 (2 c_8+c_9) B_0(s,m_1^2,m_3^2)+16 c_1 c_2 B_0(s,m_1^2,m_1^2)\nonumber\\
  &&+4 (2 c_3+7 c_4) c_2 B_0(t,m_1^2,m_2^2)+2 c_5 (2 c_8+7 c_9) B_0(t,m_1^2,m_3^2)+64 c_1 c_2 B_0(t,m_1^2,m_1^2)\nonumber\\
  &&+4 (2 c_3+c_4) c_2 B_0(u,m_1^2,m_2^2)+2 c_5 (2 c_8+c_9) B_0(u,m_1^2,m_3^2)+16 c_1 c_2 B_0(u,m_1^2,m_1^2)]\,, \nonumber\\
  %%%%
  F^{12}_{(c)}&=&2c_2+\frac{1}{16\pi^2}[4 (2 c_3+c_4) c_2 B_0(s,m_1^2,m_2^2)+2 c_5 (2 c_8+c_9) B_0(s,m_1^2,m_3^2)+16 c_1 c_2 B_0(s,m_1^2,m_1^2)\nonumber\\
  &&+4 (2 c_3+c_4) c_2 B_0(t,m_1^2,m_2^2)+2 c_5 (2 c_8+c_9) B_0(t,m_1^2,m_3^2)+16 c_1 c_2 B_0(t,m_1^2,m_1^2)\nonumber\\
  &&+4 (2 c_3+7 c_4) c_2 B_0(u,m_1^2,m_2^2)+2 c_5 (2 c_8+7 c_9) B_0(u,m_1^2,m_3^2)+64 c_1 c_2 B_0(u,m_1^2,m_1^2)]\,, \nonumber\\
  %%%%
  F^{13}_{(a)}&=&2c_5+\frac{1}{16\pi^2}[2 c_2 (2 c_8+7 c_9) B_0(s,m_1^2,m_2^2)+4 (2 c_6+7 c_7) c_5 B_0(s,m_1^2,m_3^2)+64 c_1 c_5 B_0(s,m_1^2,m_1^2)\nonumber\\
  &&+2 c_2 (2 c_8+c_9) B_0(t,m_1^2,m_2^2)+4 (2 c_6+c_7) c_5 B_0(t,m_1^2,m_3^2)+16 c_1 c_5 B_0(t,m_1^2,m_1^2)\nonumber\\
  &&+2 c_2 (2 c_8+c_9) B_0(u,m_1^2,m_2^2)+4 (2 c_6+c_7) c_5 B_0(u,m_1^2,m_3^2)+16 c_1 c_5 B_0(u,m_1^2,m_1^2)]\,, \nonumber\\
  %%%%
  F^{13}_{(b)}&=&2c_5+\frac{1}{16\pi^2}[2 c_2 (2 c_8+c_9) B_0(s,m_1^2,m_2^2)+4 (2 c_6+c_7) c_5 B_0(s,m_1^2,m_3^2)+16 c_1 c_5 B_0(s,m_1^2,m_1^2)\nonumber\\
  &&+2 c_2 (2 c_8+7 c_9) B_0(t,m_1^2,m_2^2)+4 (2 c_6+7 c_7) c_5 B_0(t,m_1^2,m_3^2)+64 c_1 c_5 B_0(t,m_1^2,m_1^2)\nonumber\\
  &&+2 c_2 (2 c_8+c_9) B_0(u,m_1^2,m_2^2)+4 (2 c_6+c_7) c_5 B_0(u,m_1^2,m_3^2)+16 c_1 c_5 B_0(u,m_1^2,m_1^2)]\,, \nonumber\\
  %%%%
  F^{13}_{(c)}&=&2c_5+\frac{1}{16\pi^2}[2 c_2 (2 c_8+c_9) B_0(s,m_1^2,m_2^2)+4 (2 c_6+c_7) c_5 B_0(s,m_1^2,m_3^2)+16 c_1 c_5 B_0(s,m_1^2,m_1^2)\nonumber\\
  &&+2 c_2 (2 c_8+c_9) B_0(t,m_1^2,m_2^2)+4 (2 c_6+c_7) c_5 B_0(t,m_1^2,m_3^2)+16 c_1 c_5 B_0(t,m_1^2,m_1^2)\nonumber\\
  &&+2 c_2 (2 c_8+7 c_9) B_0(u,m_1^2,m_2^2)+4 (2 c_6+7 c_7) c_5 B_0(u,m_1^2,m_3^2)+64 c_1 c_5 B_0(u,m_1^2,m_1^2)]\,, \nonumber\\
  %%%%
  F^{22}_{(a)}&=&2c_4+\frac{1}{16\pi^2}[8 c_4 (2 c_3+3 c_4) B_0(s,m_1^2,m_2^2)+2 c_9 (2 c_8+3 c_9) B_0(s,m_1^2,m_3^2)+16 c_2^2 B_0(s,m_1^2,m_1^2)\nonumber\\
  &&+16 c_1 c_4 B_0(t,m_1^2,m_1^2)+16 c_3 c_4 B_0(u,m_1^2,m_2^2)+4 c_8 c_9 B_0(u,m_1^2,m_3^2)+4 c_2^2 B_0(u,m_1^2,m_1^2)]\,, \nonumber\\
  %%%%
  F^{22}_{(b)}&=&4c_3+\frac{1}{16\pi^2}[4 (4 c_3^2+c_4^2) B_0(s,m_1^2,m_2^2)+(4 c_8^2+c_9^2) B_0(s,m_1^2,m_3^2)+4 c_2^2 B_0(s,m_1^2,m_1^2)\nonumber\\
  &&+16 c_1 (6 c_3+c_4) B_0(t,m_1^2,m_1^2)+4 (4 c_3^2+c_4^2) B_0(u,m_1^2,m_2^2)+(4 c_8^2+c_9^2) B_0(u,m_1^2,m_3^2)+4 c_2^2 B_0(u,m_1^2,m_1^2)]\,, \nonumber\\
  %%%%
  F^{22}_{(c)}&=&2c_4+\frac{1}{16\pi^2}[16 c_3 c_4 B_0(s,m_1^2,m_2^2)+4 c_8 c_9 B_0(s,m_1^2,m_3^2)+4 c_2^2 B_0(s,m_1^2,m_1^2)+16 c_1 c_4 B_0(t,m_1^2,m_1^2)\notag\\
  &&+8 c_4 (2 c_3+3 c_4) B_0(u,m_1^2,m_2^2)+2 c_9 (2 c_8+3 c_9) B_0(u,m_1^2,m_3^2)+16 c_2^2 B_0(u,m_1^2,m_1^2)]\,, \nonumber\\
  %%%%
  F^{23}_{(a)}&=&c_9+\frac{1}{16\pi^2}[4 (c_4 c_8+c_3 c_9+3 c_4 c_9) B_0(s,m_1^2,m_2^2)+4 (c_7 c_8+c_6 c_9+3 c_7 c_9) B_0(s,m_1^2,m_3^2)+16 c_2 c_5 B_0(s,m_1^2,m_1^2)\notag\\
  &&+8 c_1 c_9 B_0(t,m_1^2,m_1^2)+4 (c_4 c_8+c_3 c_9) B_0(u,m_1^2,m_2^2)+4 (c_7 c_8+c_6 c_9) B_0(u,m_1^2,m_3^2)+4 c_2 c_5 B_0(u,m_1^2,m_1^2)]\,, \nonumber\\
  %%%%
  F^{23}_{(b)}&=&2c_8+\frac{1}{16\pi^2}[2 (4 c_3 c_8+c_4 c_9) B_0(s,m_1^2,m_2^2)+2 (4 c_6 c_8+c_7 c_9) B_0(s,m_1^2,m_3^2)+4 c_2 c_5 B_0(s,m_1^2,m_1^2)\nonumber\\
  &&+8 c_1 (6 c_8+c_9) B_0(t,m_1^2,m_1^2)+2 (4 c_3 c_8+c_4 c_9) B_0(u,m_1^2,m_2^2)+2 (4 c_6 c_8+c_7 c_9) B_0(u,m_1^2,m_3^2)\nonumber\\
  &&+4 c_2 c_5 B_0(u,m_1^2,m_1^2)]\,, \nonumber\\
  %%%%
  F^{23}_{(c)}&=&c_9+\frac{1}{16\pi^2}[4 (c_4 c_8+c_3 c_9) B_0(s,m_1^2,m_2^2)+4 (c_7 c_8+c_6 c_9) B_0(s,m_1^2,m_3^2)+4 c_2 c_5 B_0(s,m_1^2,m_1^2)\notag\\
  &&+8 c_1 c_9 B_0(t,m_1^2,m_1^2)+4 (c_3 c_9+c_4 (c_8+3 c_9)) B_0(u,m_1^2,m_2^2)+4 (c_6 c_9+c_7 (c_8+3 c_9)) B_0(u,m_1^2,m_3^2)\notag\\
  &&+16 c_2 c_5 B_0(u,m_1^2,m_1^2)]\,, \nonumber\\
  %%%%
  F^{33}_{(a)}&=&2c_7+\frac{1}{16\pi^2}[2 c_9 (2 c_8+3 c_9) B_0(s,m_1^2,m_2^2)+8 c_7 (2 c_6+3 c_7) B_0(s,m_1^2,m_3^2)+16 c_5^2 B_0(s,m_1^2,m_1^2)\nonumber\\
  &&+16 c_1 c_7 B_0(t,m_1^2,m_1^2)+4 c_8 c_9 B_0(u,m_1^2,m_2^2)+16 c_6 c_7 B_0(u,m_1^2,m_3^2)+4 c_5^2 B_0(u,m_1^2,m_1^2)]\,, \nonumber\\
  %%%%
  F^{33}_{(b)}&=&4c_6+\frac{1}{16\pi^2}[(4 c_8^2+c_9^2) B_0(s,m_1^2,m_2^2)+4 (4 c_6^2+c_7^2) B_0(s,m_1^2,m_3^2)+4 c_5^2 B_0(s,m_1^2,m_1^2)\nonumber\\
  &&+16 c_1 (6 c_6+c_7) B_0(t,m_1^2,m_1^2)+(4 c_8^2+c_9^2) B_0(u,m_1^2,m_2^2)+4 (4 c_6^2+c_7^2) B_0(u,m_1^2,m_3^2)+4 c_5^2 B_0(u,m_1^2,m_1^2)]\,, \nonumber\\
  %%%%
  F^{33}_{(c)}&=&2c_7+\frac{1}{16\pi^2}[4 c_8 c_9 B_0(s,m_1^2,m_2^2)+16 c_6 c_7 B_0(s,m_1^2,m_3^2)+4 c_5^2 B_0(s,m_1^2,m_1^2)+16 c_1 c_7 B_0(t,m_1^2,m_1^2)\notag\\
  &&+2 c_9 (2 c_8+3 c_9) B_0(u,m_1^2,m_2^2)+8 c_7 (2 c_6+3 c_7) B_0(u,m_1^2,m_3^2)+16 c_5^2 B_0(u,m_1^2,m_1^2)]\,, 
 \end{eqnarray} 
where $s,t,u$ are the Mandelstam variables with $s=(p_1+p_3)^2$, $t=(p_1-p_3)^2$ and $u=(p_1-p_4)^2$. $m_{1,2,3}$ represent for the masses of $J/\psi$, $\psi(2S)$, and $\psi(3770)$, respectively.
Notice that the first term of each form factor is from the tree-level Feynman diagrams (LO), and the left parts are from one-loop diagrams (NLO).
The scalar function $B_0(s,m_a^2,m_b^2)$ is defined as\cite{Passarino:1978jh}
 \begin{eqnarray}
    B_0(s,m_a^2,m_b^2)&=&\frac{\lambda^{1/2}(s,m_a^2,m_b^2) }{s}\ln \left(\frac{\sqrt{s-({m_a}+{m_b})^2}-\sqrt{s-({m_a}-{m_b})^2}}{\sqrt{s-({m_a}-{m_b})^2}+\sqrt{s-({m_a}+{m_b})^2}}\right)\nonumber\\ 
    &&-\ln \left(\frac{{m_b}^2}{\mu ^2}\right)+2+\frac{\left({m_a}^2-{m_b}^2+s\right) }{2 s}\ln \left(\frac{{m_b}^2}{{m_a}^2}\right)\,,
 \end{eqnarray}
 where $\lambda(a,b,c)=(a+b-c)^2-4ab$ is the triangle function. 
 In the equal mass case, it can be simplified into 
 \begin{eqnarray}
    B_0(s,m^2,m^2)= -\ln \left(\frac{{m}^2}{\mu ^2}\right)-\rho(s,m) \ln \left(\frac{\rho(s,m)+1}{\rho(s,m)-1}\right)+2\,,
 \end{eqnarray}
with $\rho(s,m)=\sqrt{1-4m^2/s}$ the phase space factor.  Note that in the calculation of the one-loop diagrams, the propagator we used is 
 \begin{eqnarray}
     iD_{\mu\nu}(k)=-i\frac{g_{\mu\nu}-(1-\xi)\frac{k_\mu k_\nu}{k^2-m^2}}{k^2-m^2}\,,
     \label{eq:propagator}
    \end{eqnarray}
where the Feynman gauge is applied, i.e.,  $\xi=1$.
%%%
The polarization vectors are expressed as \cite{PhysRev.145.1152,Haber:1994pe}
\begin{alignat}{2}
\epsilon_1^\mu(\vec{p}_1,\pm)&=\frac{\sqrt{2}}{2}(0;\mp1,-i,0)^T &\quad\epsilon_1^\mu(\vec{p}_1,0)&=(\frac{|\vec{p}_1|}{m_1};0,0,\frac{E_1}{m_1})^T\,,\nonumber\\
%%%%%%
\epsilon_2^\mu(\vec{p}_2,\pm)&=\frac{\sqrt{2}}{2}(0;\pm1,-i,0)^T&\quad\epsilon_2^\mu(\vec{p}_2,0)&=(\frac{|\vec{p}_2|}{m_2};0,0,-\frac{E_2}{m_2})^T\,,\nonumber\\
%%%%%%
\epsilon_3^\mu(\vec{p}_3,\pm)&=\frac{\sqrt{2}}{2}(0;\mp\cos\theta_s,-i,\pm\sin\theta_s)^T&\qquad \epsilon_3^\mu(\vec{p}_3,0)&=(\frac{|\vec{p}_3|}{m_3};\frac{E_3}{m_3}\sin\theta_s,0,\frac{E_3}{m_3}\cos\theta_s)^T\,,\nonumber\\
%%%%%%
\epsilon_4^\mu(\vec{p}_4,\pm)&=\frac{\sqrt{2}}{2}(0;\pm\cos\theta_s,-i,\mp\sin\theta_s)^T&\qquad \epsilon_4^\mu(\vec{p}_4,0)&=(\frac{|\vec{p}_4|}{m_4};-\frac{E_4}{m_4}\sin\theta_s,0,-\frac{E_4}{m_4}\cos\theta_s)^T\,, \label{eq:polar;vector}
\end{alignat}
where $E_i=\sqrt{m^2+|\vec{p}_i|^2}$ is the energy of the $i-$th particle, and $\theta_s$ is the scattering angle in the $x-z$ plane. 
Here the overall phase of the polarization vector of spin-one particle has been fixed such that
\begin{eqnarray}\label{eq:pola}
\epsilon^\mu(\vec{p},-\lambda)=(-1)^\lambda\epsilon^\mu(\vec{p},\lambda)^*\,.
\end{eqnarray}
For the other particle moving in the $-\vec{p}$ direction, one has
\begin{eqnarray}\label{eq:polar3}
\epsilon^\alpha(-\vec{p},-\lambda)=-\xi_\lambda g^{\alpha\beta}\epsilon_\beta(\vec{p},\lambda)\,.
\end{eqnarray}
where $g^{\alpha\beta}={\rm diag}~(1,-1,-1,-1)$ is the metric tensor. Note that there is no summation on the indices $\alpha\alpha$. One has $\xi_\lambda=1$ for the other particle in the initial/final states in the Jacob-Wick convention \cite{Jacob:1959at}.
In summation, these spin-1 polarization vectors satisfy
\begin{eqnarray}\label{eq:polar1}
\epsilon(\vec{p},\lambda)\cdot \epsilon(\vec{p},\lambda')^{*}=-\delta_{\lambda\lambda'}\,.
\end{eqnarray}

For the scattering of the coupled channels, $J/\psi J/\psi$-$J/\psi \psi(2S)$-$J/\psi \psi(3770)$, the 41 independent helicity amplitudes are given as:
\begin{eqnarray}\label{eq:expand polar}
  T^{ij}_{++++}(s,z_s)&=&F^{ij}_{(a)}+\frac{1}{4} (z_s+1)^2F^{ij}_{(b)}+\frac{1}{4} (z_s-1)^2F^{ij}_{(c)}\,, \nonumber\\
  T^{ij}_{+++-}(s,z_s)&=&\frac{1}{4} \left(1-z_s^2\right)(F^{ij}_{(b)}+F^{ij}_{(c)})\,, \nonumber\\
 T^{ij}_{+-++}(s,z_s)&=&\frac{1}{4} \left(1-z_s^2\right)(F^{ij}_{(b)}+F^{ij}_{(c)})\,, \nonumber\\
 T^{ij}_{++-+}(s,z_s)&=&\frac{1}{4} \left(1-z_s^2\right)(F^{ij}_{(b)}+F^{ij}_{(c)})\,, \nonumber\\
 T^{ij}_{-+++}(s,z_s)&=&\frac{1}{4} \left(1-z_s^2\right)(F^{ij}_{(b)}+F^{ij}_{(c)})\,, \nonumber\\
 T^{ij}_{+++0}(s,z_s)&=&-(\frac{\sqrt{2}(z_s+1) \sqrt{1-z_s^2}  {E_2}}{4 {m_2}}F^{ij}_{(b)}+\frac{\sqrt{2}(z_s-1) \sqrt{1-z_s^2}  {E_2}}{4 {m_2}}F^{ij}_{(c)})\,, \nonumber\\
 T^{ij}_{+0++}(s,z_s)&=&(\frac{\sqrt{2}(z_s+1) \sqrt{1-z_s^2}  {E_4}}{4 {m_4}}F^{ij}_{(b)}+\frac{\sqrt{2}(z_s-1) \sqrt{1-z_s^2}  {E_4}}{4 {m_4}}F^{ij}_{(c)})\,, \nonumber\\
 T^{ij}_{++0+}(s,z_s)&=&\frac{\sqrt{2}(z_s+1) \sqrt{1-z_s^2}  {E_1}}{4 {m_1}}F^{ij}_{(b)}+\frac{\sqrt{2}(z_s-1) \sqrt{1-z_s^2}  {E_1}}{4 {m_1}}F^{ij}_{(c)}\,, \nonumber\\
 T^{ij}_{0+++}(s,z_s)&=&-\frac{\sqrt{2}(z_s+1) \sqrt{1-z_s^2}  {E_3}}{4 {m_3}}F^{ij}_{(b)}-\frac{\sqrt{2}(z_s-1) \sqrt{1-z_s^2}  {E_3}}{4 {m_3}}F^{ij}_{(c)}\,, \nonumber\\
 T^{ij}_{++--}(s,z_s)&=&F^{ij}_{(a)}+\frac{1}{4} (z_s-1)^2F^{ij}_{(b)}+\frac{1}{4} (z_s+1)^2F^{ij}_{(c)}\,, \nonumber\\
 T^{ij}_{+-+-}(s,z_s)&=&\frac{1}{4} (z_s+1)^2(F^{ij}_{(b)}+F^{ij}_{(c)})\,, \nonumber\\
 T^{ij}_{+--+}(s,z_s)&=&\frac{1}{4} (z_s-1)^2(F^{ij}_{(b)}+F^{ij}_{(c)})\,, \nonumber\\
 T^{ij}_{++-0}(s,z_s)&=&\frac{\sqrt{2} (z_s-1) \sqrt{1-z_s^2} {E_2}}{4 {m_2}}F^{ij}_{(b)}+\frac{\sqrt{2} (z_s+1) \sqrt{1-z_s^2} {E_2}}{4 {m_2}}F^{ij}_{(c)}\,, \nonumber\\
 T^{ij}_{-0++}(s,z_s)&=&-\frac{\sqrt{2} (z_s-1) \sqrt{1-z_s^2} {E_4}}{4 {m_4}}F^{ij}_{(b)}-\frac{\sqrt{2} (z_s+1) \sqrt{1-z_s^2} {E_4}}{4 {m_4}}F^{ij}_{(c)}\,, \nonumber\\
 T^{ij}_{++0-}(s,z_s)&=&\frac{\sqrt{2} (1-z_s) \sqrt{1-z_s^2} {E_1}}{4 {m_1}}F^{ij}_{(b)}+\frac{\sqrt{2} (-1-z_s) \sqrt{1-z_s^2} {E_1}}{4 {m_1}}F^{ij}_{(c)}\,, \nonumber\\
 T^{ij}_{0-++}(s,z_s)&=&-\frac{\sqrt{2} (1-z_s) \sqrt{1-z_s^2} {E_3}}{4 {m_3}}F^{ij}_{(b)}-\frac{\sqrt{2} (-1-z_s) \sqrt{1-z_s^2} {E_3}}{4 {m_3}}F^{ij}_{(c)}\,, \nonumber\\
 T^{ij}_{+-+0}(s,z_s)&=&\frac{\sqrt{2} (z_s+1) \sqrt{1-z_s^2} {E_2}}{4 {m_2}}(F^{ij}_{(b)}+F^{ij}_{(c)})\,, \nonumber\\
 T^{ij}_{+0+-}(s,z_s)&=&-\frac{\sqrt{2} (z_s+1) \sqrt{1-z_s^2} {E_4}}{4 {m_4}}(F^{ij}_{(b)}+F^{ij}_{(c)})\,, \nonumber\\
  T^{ij}_{+-0+}(s,z_s)&=&\frac{\sqrt{2} (1-z_s) \sqrt{1-z_s^2} {E_1}}{4 {m_1}}(F^{ij}_{(b)}+F^{ij}_{(c)})\,, \nonumber\\
  T^{ij}_{0++-}(s,z_s)&=&-\frac{\sqrt{2} (1-z_s) \sqrt{1-z_s^2} {E_3}}{4 {m_3}}(F^{ij}_{(b)}+F^{ij}_{(c)})\,, \nonumber\\
  T^{ij}_{-++0}(s,z_s)&=&-\frac{\sqrt{2} (1-z_s) \sqrt{1-z_s^2} {E_2}}{4 {m_2}}(F^{ij}_{(b)}+F^{ij}_{(c)})\,, \nonumber\\
  T^{ij}_{+0-+}(s,z_s)&=&\frac{\sqrt{2} (1-z_s) \sqrt{1-z_s^2} {E_4}}{4 {m_4}}(F^{ij}_{(b)}+F^{ij}_{(c)})\,, \nonumber\\
 T^{ij}_{-+0+}(s,z_s)&=&\frac{\sqrt{2} (-z_s-1) \sqrt{1-z_s^2} {E_1}}{4 {m_1}}(F^{ij}_{(b)}+F^{ij}_{(c)})\,, \nonumber\\
 T^{ij}_{0+-+}(s,z_s)&=&-\frac{\sqrt{2} (-z_s-1) \sqrt{1-z_s^2} {E_3}}{4 {m_3}}(F^{ij}_{(b)}+F^{ij}_{(c)})\,, \nonumber\\
 T^{ij}_{++00}(s,z_s)&=&-\frac{{E_1E_2}+{p_{cm}^2}}{{m_1} {m_2}}F^{ij}_{(a)}+\frac{\left(z_s^2-1\right) {E_1E_2}}{2 {m_1} {m_2}}(F^{ij}_{(b)}+F^{ij}_{(c)})\,, \nonumber\\
 T^{ij}_{00++}(s,z_s)&=&-\frac{{E_3E_4}+(p_{cm}^{\prime})^2}{{m_3} {m_4}}F^{ij}_{(a)}+\frac{\left(z_s^2-1\right) {E_3E_4}}{2 {m_3} {m_4}}(F^{ij}_{(b)}+F^{ij}_{(c)})\,, \nonumber\\
 T^{ij}_{+0+0}(s,z_s)&=&\frac{(z_s+1) (z_s {E_2E_4}-p_{cm}p_{cm}^\prime)}{2 {m_2} {m_4}}F^{ij}_{(b)}+\frac{\left(z_s^2-1\right) {E_2E_4}}{2 {m_2} {m_4}}F^{ij}_{(c)}\,, \nonumber\\
 T^{ij}_{+00+}(s,z_s)&=&-(\frac{\left(z_s^2-1\right) {E_1E_4}}{2 {m_1} {m_4}}F^{ij}_{(b)}+\frac{(z_s-1) (z_s {E_1E_4}+p_{cm}p_{cm}^\prime)}{2 {m_1} {m_4}}F^{ij}_{(c)})\,, \nonumber\\
 T^{ij}_{0++0}(s,z_s)&=&-(\frac{\left(z_s^2-1\right) {E_2E_3}}{2 {m_2} {m_3}}F^{ij}_{(b)}+\frac{(z_s-1) (z_s {E_2E_3}+p_{cm}p_{cm}^\prime)}{2 {m_2} {m_3}}F^{ij}_{(c)})\,, \nonumber\\
 T^{ij}_{0+0+}(s,z_s)&=&\frac{(z_s+1) (z_s {E_1E_3}-p_{cm}p_{cm}^\prime)}{2 {m_1} {m_3}}F^{ij}_{(b)}+\frac{\left(z_s^2-1\right) {E_1E_3}}{2 {m_1} {m_3}}F^{ij}_{(c)}\,, \nonumber\\
 T^{ij}_{+0-0}(s,z_s)&=&\frac{(z_s-1) (p_{cm}p_{cm}^\prime-z_s {E_2E_4})}{2 {m_2} {m_4}}F^{ij}_{(b)}+\frac{\left(1-z_s^2\right) {E_2E_4}}{2 {m_2} {m_4}}F^{ij}_{(c)}\,, \nonumber\\
 T^{ij}_{+00-}(s,z_s)&=&-(\frac{\left(1-z_s^2\right) {E_1E_4}}{2 {m_1} {m_4}}F^{ij}_{(b)}+\frac{(z_s+1) (z_s (-{E_1E_4})-p_{cm}p_{cm}^\prime)}{2 {m_1} {m_4}}F^{ij}_{(c)})\,, \nonumber\\
 T^{ij}_{0-+0}(s,z_s)&=&-(\frac{\left(1-z_s^2\right) {E_2E_3}}{2 {m_2} {m_3}}F^{ij}_{(b)}+\frac{(z_s+1) (z_s (-{E_2E_3})-p_{cm}p_{cm}^\prime)}{2 {m_2} {m_3}}F^{ij}_{(c)})\,, \nonumber\\
 T^{ij}_{0+0-}(s,z_s)&=&\frac{(z_s-1) (p_{cm}p_{cm}^\prime-z_s {E_1E_3})}{2 {m_1} {m_3}}F^{ij}_{(b)}+\frac{\left(1-z_s^2\right) {E_1E_3}}{2 {m_1} {m_3}}F^{ij}_{(c)}\,, \nonumber\\
 T^{ij}_{+-00}(s,z_s)&=&\frac{\left(1-z_s^2\right) {E_1E_2}}{2 {m_1} {m_2}}(F^{ij}_{(b)}+F^{ij}_{(c)})\,, \nonumber\\
 T^{ij}_{00+-}(s,z_s)&=&\frac{\left(1-z_s^2\right) {E_3E_4}}{2 {m_3} {m_4}}(F^{ij}_{(b)}+F^{ij}_{(c)})\,, \nonumber\\
 T^{ij}_{+000}(s,z_s)&=&\frac{\sqrt{2} \sqrt{1-z_s^2} {E_1} (z_s {E_2E_4}-p_{cm}p_{cm}^\prime)}{2 {m_1} {m_2} {m_4}}F^{ij}_{(b)}+\frac{\sqrt{2} \sqrt{1-z_s^2} {E_2} (z_s {E_1E_4}+p_{cm}p_{cm}^\prime)}{2 {m_1} {m_2} {m_4}}F^{ij}_{(c)}\,, \nonumber\\
 T^{ij}_{00+0}(s,z_s)&=&-\frac{\sqrt{2} \sqrt{1-z_s^2} {E_3} (z_s {E_2E_4}-p_{cm}p_{cm}^\prime)}{2 {m_2} {m_3} {m_4}}F^{ij}_{(b)}-\frac{\sqrt{2} \sqrt{1-z_s^2} {E_4} (z_s {E_2E_3}+p_{cm}p_{cm}^\prime)}{2 {m_2} {m_3} {m_4}}F^{ij}_{(c)}\,, \nonumber\\
T^{ij}_{0+00}(s,z_s)&=&\frac{\sqrt{2} \sqrt{1-z_s^2} {E_2} (-z_s {E_1E_3}+p_{cm}p_{cm}^\prime)}{2 {m_1} {m_2} {m_3}}F^{ij}_{(b)}-\frac{\sqrt{2} \sqrt{1-z_s^2} {E_1} (z_s {E_2E_3}+p_{cm}p_{cm}^\prime)}{2 {m_1} {m_2} {m_3}}F^{ij}_{(c)}\,, \nonumber\\
T^{ij}_{000+}(s,z_s)&=&-\frac{\sqrt{2} \sqrt{1-z_s^2} {E_4} (-z_s {E_1E_3}+p_{cm}p_{cm}^\prime)}{2 {m_1} {m_3} {m_4}}F^{ij}_{(b)}+\frac{\sqrt{2} \sqrt{1-z_s^2} {E_3} (z_s {E_1E_4}+p_{cm}p_{cm}^\prime)}{2 {m_1} {m_3} {m_4}}F^{ij}_{(c)}\,, \nonumber\\
 T^{ij}_{0000}(s,z_s)&=&\frac{({E_1E_2}+{p_{cm}^2) ({E_3E_4}+(p_{cm}^{\prime})^2)}}{{m_1} {m_2} {m_3} {m_4}}F^{ij}_{(a)}+\frac{({p_{cm}p_{cm}^\prime-z_s {E_1E_3}}) ({p_{cm}p_{cm}^\prime-z_s {E_2E_4})}}{{m_1} {m_2} {m_3} {m_4}}F^{ij}_{(b)}+\notag\\
  &&\frac{(z_s {E_1E_4}+{p_{cm}p_{cm}^\prime}) (z_s {E_2E_3}+{p_{cm}p_{cm}^\prime})}{{m_1} {m_2} {m_3} {m_4}}F^{ij}_{(c)}\,, 
\end{eqnarray}
where  $E_i=\sqrt{m_i^2+|\vec{p}_i|^2}$ is the energy of the  $i$-th particle (i=1, 2, 3, 4) in the center of mass frame, and $|\vec{p}_i|$ is the modulus of the three momentum. One has $|\vec{p}_1|=|\vec{p}_2|=p_{cm}=\sqrt{\lambda(s,m_1^2,m_2^2)/4s}$, $|\vec{p}_3|=|\vec{p}_4|=p_{cm}^\prime=\sqrt{\lambda(s,m_3^2,m_4^2)/4s}$.

The decompositions of the helicity amplitudes in $|JMLS\rangle$ representation are given as:
\begin{eqnarray}
   T^{ij}_{\mu_1\mu_2;\mu_3\mu_4}(s,z_s) &=& 16\pi N_{ij}\sum_J (2J+1) d_{\mu\mu'}^J(z_s) \sum_{LS,L'S'}\frac{\sqrt{(2L+1)(2L'+1)}}{2J+1}\left\langle LS0\mu|J\mu\right\rangle \left\langle J\mu'| L'S'0\mu'\right\rangle \nonumber\\
  && \left\langle s_1s_2\mu_1,-\mu_2|S\mu\right\rangle \left\langle S'\mu' |s_3s_4\mu_3,-\mu_4 \right\rangle 
  T^{J,ij}_{LS,L'S'}  \,. \label{eq:pwLS}
\end{eqnarray}
\begin{eqnarray}
  T^{ij}_{++++}(s,z_s)&=&\frac{16\pi}{3}T^{ij}_{^1S_0}(s)+8\pi T^{ij}_{^3P_0}(s)+[\frac{4\pi}{3}T^{ij}_{^5S_2}(s)+8\pi T^{ij}_{^3P_2}(s)](3z_s^2-1)+\cdots\,, \nonumber\\
  T^{ij}_{+++-}(s,z_s)&=&4\pi T^{ij}_{^5S_2}(s)(1-z_s^2)+\cdots\,, \nonumber\\
   T^{ij}_{+-++}(s,z_s)&=&4\pi T^{ij}_{^5S_2}(s)(1-z_s^2)+\cdots\,, \nonumber\\
   T^{ij}_{++-+}(s,z_s)&=&4\pi T^{ij}_{^5S_2}(s)(1-z_s^2)+\cdots\,, \nonumber\\
   T^{ij}_{-+++}(s,z_s)&=&4\pi T^{ij}_{^5S_2}(s)(1-z_s^2)+\cdots\,, \nonumber\\
   T^{ij}_{+++0}(s,z_s)&=&[4\sqrt{2}\pi T^{ij}_{^5S_2}(s)+12\sqrt{2}\pi T^{ij}_{^3P_2}(s)]\sqrt{1-z_s^2}z_s+\cdots\,, \nonumber\\
   T^{ij}_{+0++}(s,z_s)&=&[-4\sqrt{2}\pi T^{ij}_{^5S_2}(s)-12\sqrt{2}\pi T^{ij}_{^3P_2}(s)]\sqrt{1-z_s^2}z_s+\cdots\,, \nonumber\\
   T^{ij}_{++0+}(s,z_s)&=&[-4\sqrt{2}\pi T^{ij}_{^5S_2}(s)-12\sqrt{2}\pi T^{ij}_{^3P_2}(s)]\sqrt{1-z_s^2}z_s+\cdots\,, \nonumber\\
   T^{ij}_{0+++}(s,z_s)&=&[4\sqrt{2}\pi T^{ij}_{^5S_2}(s)+12\sqrt{2}\pi T^{ij}_{^3P_2}(s)]\sqrt{1-z_s^2}z_s+\cdots\,, \nonumber\\
   T^{ij}_{++--}(s,z_s)&=&\frac{16\pi}{3}T^{ij}_{^1S_0}(s)-8\pi T^{ij}_{^3P_0}(s)+[\frac{4\pi}{3} T^{ij}_{^5S_2}(s)-8\pi T^{ij}_{^3P_2}(s)](3z_s^2-1)+\cdots\,, \nonumber\\
   T^{ij}_{+-+-}(s,z_s)&=&4\pi T^{ij}_{^5S_2}(s)(1+z_s)^2+\cdots\,, \nonumber\\
   T^{ij}_{+--+}(s,z_s)&=&4\pi T^{ij}_{^5S_2}(s)(1-z_s)^2+\cdots\,, \nonumber\\
   T^{ij}_{++-0}(s,z_s)&=&[-4\sqrt{2}\pi T^{ij}_{^5S_2}(s)+12\sqrt{2}\pi T^{ij}_{^3P_2}(s)]\sqrt{1-z_s^2}z_s+\cdots\,, \nonumber\\
   T^{ij}_{-0++}(s,z_s)&=&(4\sqrt{2}\pi T^{ij}_{^5S_2}(s)-12\sqrt{2}\pi T^{ij}_{^3P_2}(s))\sqrt{1-z_s^2}z_s+\cdots\,, \nonumber\\
   T^{ij}_{++0-}(s,z_s)&=&[4\sqrt{2}\pi T^{ij}_{^5S_2}(s)-12\sqrt{2}\pi T^{ij}_{^3P_2}(s)]\sqrt{1-z_s^2}z_s+\cdots\,, \nonumber\\
   T^{ij}_{0-++}(s,z_s)&=&[-4\sqrt{2}\pi T^{ij}_{^5S_2}(s)+12\sqrt{2}\pi T^{ij}_{^3P_2}(s)]\sqrt{1-z_s^2}z_s+\cdots\,, \nonumber\\
   T^{ij}_{+-+0}(s,z_s)&=&-4\sqrt{2}\pi T^{ij}_{^5S_2}(s)(1+z_s)\sqrt{1-z_s^2}+\cdots\,, \nonumber\\
   T^{ij}_{+0+-}(s,z_s)&=&4\sqrt{2}\pi T^{ij}_{^5S_2}(s)(1+z_s)\sqrt{1-z_s^2}+\cdots\,, \nonumber\\
   T^{ij}_{+-0+}(s,z_s)&=&-4\sqrt{2}\pi T^{ij}_{^5S_2}(s)(1-z_s)\sqrt{1-z_s^2}+\cdots\,, \nonumber\\
   T^{ij}_{0++-}(s,z_s)&=&4\sqrt{2}\pi T^{ij}_{^5S_2}(s)(1-z_s)\sqrt{1-z_s^2}+\cdots\,, \nonumber\\
   T^{ij}_{-++0}(s,z_s)&=&-4\sqrt{2}\pi T^{ij}_{^5S_2}(s)(z_s-1)\sqrt{1-z_s^2}+\cdots\,, \nonumber\\
   T^{ij}_{+0-+}(s,z_s)&=&4\sqrt{2}\pi T^{ij}_{^5S_2}(s)(z_s-1)\sqrt{1-z_s^2}+\cdots\,, \nonumber\\
   T^{ij}_{-+0+}(s,z_s)&=&4\sqrt{2}\pi T^{ij}_{^5S_2}(s)(z_s+1)\sqrt{1-z_s^2}+\cdots\,, \nonumber\\
   T^{ij}_{0+-+}(s,z_s)&=&-4\sqrt{2}\pi T^{ij}_{^5S_2}(s)(z_s+1)\sqrt{1-z_s^2}+\cdots\,, \nonumber\\
   T^{ij}_{++00}(s,z_s)&=&-\frac{16\pi}{3}T^{ij}_{^1S_0}(s)+\frac{8\pi}{3}T^{ij}_{^5S_2}(s)(3z_s^2-1)+\cdots\,, \nonumber\\
   T^{ij}_{00++}(s,z_s)&=&-\frac{16\pi}{3}T^{ij}_{^1S_0}(s)+\frac{8\pi}{3}T^{ij}_{^5S_2}(s)(3z_s^2-1)+\cdots\,, \nonumber\\
   T^{ij}_{+0+0}(s,z_s)&=&6\pi T^{ij}_{^3P_1}(s)(1+z_s)+[4\pi T^{ij}_{^5S_2}(s)+6\pi T^{ij}_{^3P_2}(s)](2z_s^2+z_s-1)+\cdots\,, \nonumber\\
   T^{ij}_{+00+}(s,z_s)&=&-6\pi T^{ij}_{^3P_1}(s)(1-z_s)+[4\pi T^{ij}_{^5S_2}(s)+6\pi T^{ij}_{^3P_2}(s)](-2z_s^2+z_s+1)+\cdots\,, \nonumber\\
   T^{ij}_{0++0}(s,z_s)&=&-6\pi T^{ij}_{^3P_1}(s)(1-z_s)+[4\pi T^{ij}_{^5S_2}(s)+6\pi T^{ij}_{^3P_2}(s)](-2z_s^2+z_s+1)+\cdots\,, \nonumber\\
   T^{ij}_{0+0+}(s,z_s)&=&6\pi T^{ij}_{^3P_1}(s)(1+z_s)+[4\pi T^{ij}_{^5S_2}(s)+6\pi T^{ij}_{^3P_2}(s)](2z_s^2+z_s-1)+\cdots\,, \nonumber\\
   T^{ij}_{+0-0}(s,z_s)&=&6\pi T^{ij}_{^3P_1}(s)(1-z_s)+[4\pi T^{ij}_{^5S_2}(s)-6\pi T^{ij}_{^3P_2}(s)](-2z_s^2+z_s+1)+\cdots\,, \nonumber\\
   T^{ij}_{+00-}(s,z_s)&=&-6\pi T^{ij}_{^3P_1}(s)(1+z_s)+[4\pi T^{ij}_{^5S_2}(s)-6\pi T^{ij}_{^3P_2}(s)](2z_s^2+z_s-1)+\cdots\,, \nonumber\\
   T^{ij}_{0-+0}(s,z_s)&=&-6\pi T^{ij}_{^3P_1}(s)(1+z_s)+[4\pi T^{ij}_{^5S_2}(s)-6\pi T^{ij}_{^3P_2}(s)](2z_s^2+z_s-1)+\cdots\,, \nonumber\\
   T^{ij}_{0+0-}(s,z_s)&=&6\pi T^{ij}_{^3P_1}(s)(1-z_s)+[4\pi T^{ij}_{^5S_2}(s)-6\pi T^{ij}_{^3P_2}(s)](-2z_s^2+z_s+1)+\cdots\,, \nonumber\\
   T^{ij}_{+-00}(s,z_s)&=&8\pi T^{ij}_{^5S_2}(s)(1-z_s^2)+\cdots\,, \nonumber\\
   T^{ij}_{00+-}(s,z_s)&=&8\pi T^{ij}_{^5S_2}(s)(1-z_s^2)+\cdots\,, \nonumber\\
   T^{ij}_{+000}(s,z_s)&=&-8\sqrt{2}\pi T^{ij}_{^5S_2}(s)\sqrt{1-z_s^2}z_s+\cdots\,, \nonumber\\
   T^{ij}_{00+0}(s,z_s)&=&8\sqrt{2}\pi T^{ij}_{^5S_2}(s)\sqrt{1-z_s^2}z_s+\cdots\,, \nonumber\\
   T^{ij}_{0+00}(s,z_s)&=&8\sqrt{2}\pi T^{ij}_{^5S_2}(s)\sqrt{1-z_s^2}z_s+\cdots\,, \nonumber\\
   T^{ij}_{000+}(s,z_s)&=&-8\sqrt{2}\pi T^{ij}_{^5S_2}(s)\sqrt{1-z_s^2}z_s+\cdots\,, \nonumber\\
   T^{ij}_{0000}(s,z_s)&=&\frac{16\pi}{3}T^{ij}_{^1S_0}(s)+\frac{16\pi}{3}T^{ij}_{^5S_2}(s)(3z_s^2-1)+\cdots\,, \label{eq:Tpw;A}
\end{eqnarray}
where the ellipses represent the ignored higher partial waves. With it, the summation of the square of the helicity amplitudes can be obtained. See Eq.~(\ref{eq:mod;T}). 
According to Eqs.(\ref{eq:pwJ},\ref{eq:pw41},\ref{eq:pw25},\ref{eq:expand polar}), the partial wave amplitudes can be obtained as follows:
\begin{eqnarray}\label{eq:pw amp}
 T^{ii}_{^1S_0}&=&\frac{1}{32\pi N_{ii}}\int_{-1}^1\frac{1}{3m_1m_2m_3m_4}\{F^{ii}_{(a)}[(E_1 E_2+p^2) (E_3 E_4+(p_{cm}^\prime)^2)+4 m_3 m_4 (E_1 E_2+m_1 m_2+p_{cm}^2)]\notag\\
  &&+F^{ii}_{(b)}[(p_{cm} p_{cm}^\prime-E_1 E_3 z) (p_{cm} p_{cm}^\prime-E_2 E_4 z)+m_3 m_4 (m_1 m_2 (z_s^2+1)-2 E_1 E_2 (z_s^2-1))]\notag\\
  &&+F^{ii}_{(c)}[(E_1 E_4 z+p_{cm} p_{cm}^\prime) (E_2 E_3 z+p_{cm} p_{cm}^\prime)+m_3 m_4 (m_1 m_2 (z_s^2+1)-2 E_1 E_2 (z_s^2-1))]\}dz_s\,, \nonumber\\
  %%%%%
   T^{ii}_{^5S_2}&=&\frac{1}{32\pi N_{ii}}\int_{-1}^1\frac{1}{30m_1m_2m_3m_4}\{F^{ii}_{(a)}[2 (3 z_s^2-1) ((E_1 E_2+p_{cm}^2) (E_3 E_4+(p_{cm}^\prime)^2)+m_3 m_4 (m_1 m_2-2 (E_1 E_2\notag\\
   &&+p_{cm}^2)))]+F^{ii}_{(b)}[2 ((p_{cm} p_{cm}^\prime-E_1 E_3 z_s) ((3 z_s^2-1) (p_{cm} p_{cm}^\prime-E_2 E_4 z)+m_4 (6 E_2 z_s (z_s^2-1)-3 m_2 z_s^3))\nonumber\\
   &&+m_3 (m_1 (3 z_s^3 (E_2 E_4 z-p_{cm} p_{cm}^\prime)+m_4 (3 E_2 (-2 z_s^4+z_s^2+1)+m_2 (3 z_s^4+2 z_s^2+2)))-E_1 (z_s^2-1) (6 z (E_2 E_4 z_s\nonumber\\
   &&-p_{cm} p_{cm}^\prime)+E_2 m_4 (4-6 z_s^2)+3 m_2 (-2 E_4 z_s^2+E_4+m_4 (2 z_s^2+1)))))]+F^{ii}_{(c)}[2 (3 z_s^2-1) (E_1 E_4 z+p_{cm} p_{cm}^\prime)*\nonumber\\
   && (E_2 E_3 z_s+p_{cm} p_{cm}^\prime)-3 E_1 m_4 (z_s^2-1) (4 z_s (E_2 E_3 z_s+p_{cm} p_{cm}^\prime)+m_2 (E_3-2 E_3 z_s^2))+m_3 (-4 E_2 (z_s^2-1)*\nonumber\\
   && (3 z_s (E_1 E_4 z_s+p_{cm} p_{cm}^\prime)+E_1 m_4 (2-3 z_s^2))+6 m_2 (2 z_s^3 (E_1 E_4 z+p_{cm} p_{cm}^\prime)+E_1 m_4 (-2 z_s^4+z_s^2+1))\nonumber\\
   &&+m_1 (3 E_2 E_4 (2 z_s^4-3 z_s^2+1)+m_4 (6 E_2 (-2 z_s^4+z_s^2+1)+m_2 (6 z_s^4+4 z_s^2+4))))]\}dz_s\,,\nonumber\\
 %%%%%%
   T^{ii}_{^3P_0}&=&\frac{1}{32\pi N_{ii}}\int_{-1}^1(F^{ii}_{(b)}-F^{ii}_{(c)})z_sdz_s\,, \nonumber\\
   %%%%%%
   T^{ii}_{^3P_1}&=&\frac{1}{32\pi N_{ii}}\int_{-1}^1\frac{1}{4m_1m_2m_3m_4}\{F^{ii}_{(b)}[m_2 (-m_4 (z_s^2+1) (p_{cm} p_{cm}^\prime-E_1 E_3 z_s)-2 E_1 E_4 m_3 z_s (z_s^2-1))\notag\\
   &&-m_1 m_3 (z_s^2+1) (p_{cm} p_{cm}^\prime-E_2 E_4 z)]+F^{ii}_{(c)}[m_2 (E_1 E_3 m_4 z (z_s^2-1)-2 m_3 (z_s^2+1) (E_1 E_4 z_s+p_{cm} p_{cm}^\prime))\notag\\
   &&+E_2 E_4 m_1 m_3 z_s (z_s^2-1)]\}dz_s\,, \nonumber\\
   %%%%%%
   T^{ii}_{^3P_2}&=&\frac{1}{32\pi N_{ii}}\int_{-1}^1\frac{1}{20m_1m_2m_3m_4}\{F^{ii}_{(b)}[3 m_2 (m_4 (3 z_s^2-1) (-(p_{cm} p_{cm}^\prime-E_1 E_3 z_s))-2 E_1 m_3 z_s (z_s^2-1) (E_4+2 m_4))\nonumber\\
   &&+m_1 m_3 (-3 (3 z_s^2-1) (p_{cm} p_{cm}^\prime-E_2 E_4 z)-4 m_4 z_s (3 E_2 (z_s^2-1)+m_2 (1-3 z_s^2)))]\nonumber\\
   &&+F^{ii}_{(c)}[3 m_2 (E_1 E_3 m_4 z_s (z_s^2-1)+m_3 (4 E_1 m_4 z_s (z_s^2-1)-2 (3 z_s^2-1) (E_1 E_4 z_s+p_{cm} p_{cm}^\prime)))\nonumber\\
   &&+m_1 m_3 z_s (3 E_2 E_4 (z_s^2-1)+4 m_4 (3 E_2 (z_s^2-1)+m_2 (1-3 z_s^2)))]\}dz_s\,, \nonumber\\
   %%%%%%%%%%%%%%%%%%%%%%%%%%
   T^{ij}_{^1S_0}&=&\frac{1}{32\pi N_{ij}}\int_{-1}^1\frac{1}{3m_1m_2m_3m_4}\{F^{ij}_{(a)}[(E_1 E_2+2 m_1 m_2+p^2) (E_3 E_4+2 m_3 m_4+(p_{cm})^2)]+F^{ij}_{(b)}[(p_{cm} p_{cm}-E_1 E_3 z_s) \notag\\
   &&*(p_{cm} p_{cm}-E_2 E_4 z_s)-E_1 E_2 m_3 m_4 (z_s^2-1)+m_1 m_2 (m_3 m_4 (z_s^2+1)-E_3 E_4 (z_s^2-1))]+F^{ij}_{(c)}[(E_1 E_4 z_s\notag\\
   &&+p_{cm} p_{cm}) (E_2 E_3 z_s+p_{cm} p_{cm})-E_1 E_2 m_3 m_4 (z_s^2-1)+m_1 m_2 (m_3 m_4 (z_s^2+1)-E_3 E_4 (z_s^2-1))]\}dz_s\,, \nonumber\\
   %%%%%
   T^{ij}_{^5S_2}&=&\frac{1}{32\pi N_{ij}}\int_{-1}^1\frac{1}{30m_1m_2m_3m_4}\{F^{ij}_{(a)}[2 (3 z_s^2-1) (E_1 E_2-m_1 m_2+p_{cm}^2) (E_3 E_4-m_3 m_4+(p_{cm}^\prime)^2)]+F^{ij}_{(b)}[2 (p_{cm} p_{cm}^\prime\notag\\
   &&-E_2 E_4 z_s) ((3 z_s^2-1) (p_{cm} p_{cm}^\prime-E_1 E_3 z_s)+3 E_1 m_3 z_s (z_s^2-1))+2 E_2 m_4 (z_s^2-1) (3 z_s (p_{cm} p_{cm}^\prime-E_1 E_3 z_s)\nonumber\\
   &&+E_1 m_3 (3 z_s^2-2))+3 m_2 (2 (E_4 z_s (z_s^2-1)-m_4 z_s^3) (p_{cm} p_{cm}^\prime-E_1 E_3 z_s)+E_1 m_3 (z_s^2-1) (E_4 (2 z_s^2-1)\nonumber\\
   &&-m_4 (2 z_s^2+1)))+m_1 (6 z_s (E_3 (z_s^2-1)-m_3 z_s^2) (p_{cm} p_{cm}^\prime-E_2 E_4 z_s)+3 E_2 m_4 (z_s^2-1) (E_3 (2 z_s^2-1)\nonumber\\
   &&-m_3 (2 z_s^2+1))+m_2 (E_4 (z_s^2-1) (2 E_3 (3 z_s^2-2)-3 m_3 (2 z_s^2+1))+m_4 (3 E_3 (-2 z_s^4+z_s^2+1)\nonumber\\
   &&+m_3 (6 z_s^4+4 z_s^2+4))))]+F^{ij}_{(c)}[6 (m_2 z_s (-E_3 z_s^2+E_3+m_3 z_s^2)-E_2 m_3 z_s (z_s^2-1)) (E_1 E_4 z_s+p_{cm} p_{cm}^\prime)\nonumber\\
   &&+2 (3 z_s^2-1) (E_1 E_4 z_s+p_{cm} p_{cm}^\prime) (E_2 E_3 z_s+p_{cm} p_{cm}^\prime)-E_1 m_4 (z_s^2-1) (6 z_s (E_2 E_3 z_s+p_{cm} p_{cm}^\prime)\nonumber\\
   &&+2 E_2 m_3 (2-3 z_s^2)+3 m_2 (-2 E_3 z_s^2+E_3+2 m_3 z_s^2+m_3))+m_1 (m_4 (6 z_s^3 (E_2 E_3 z_s+p_{cm} p_{cm}^\prime)\nonumber\\
   &&+m_3 (3 E_2 (-2 z_s^4+z_s^2+1)+m_2 (6 z_s^4+4 z_s^2+4))+3 E_3 m_2 (-2 z_s^4+z_s^2+1))-E_4 (z_s^2-1) (6 z_s (E_2 E_3 z_s\nonumber\\
   &&+p_{cm} p_{cm}^\prime)+3 E_2 m_3 (1-2 z_s^2)+m_2 (-6 E_3 z_s^2+4 E_3+m_3 (6 z_s^2+3))))]\}dz_s\,,\nonumber\\
 %%%%%%
   T^{ij}_{^3P_0}&=&\frac{1}{32\pi N_{ij}}\int_{-1}^1(F^{ij}_{(b)}-F^{ij}_{(c)})z_sdz_s\,, \nonumber\\
   %%%%%%
   T^{ij}_{^3P_1}&=&\frac{1}{32\pi N_{ij}}\int_{-1}^1\frac{1}{4m_1m_2m_3m_4}\{F^{ij}_{(b)}[m_2 (m_4 (z_s^2+1) (E_1 E_3 z_s-p_{cm} p_{cm})-E_1 E_4 m_3 z_s (z_s^2-1))+m_1 (m_3 (z_s^2+1) *\nonumber\\
   &&(-(p_{cm} p_{cm}-E_2 E_4 z_s))-E_2 E_3 m_4 z_s (z_s^2-1))]+F^{ij}_{(c)}[m_2 (E_1 E_3 m_4 z_s (z_s^2-1)-m_3 (z_s^2+1) (E_1 E_4 z_s\nonumber\\
   &&+p_{cm} p_{cm}))+m_1 (E_2 E_4 m_3 z_s (z_s^2-1)-m_4 (z_s^2+1) (E_2 E_3 z_s+p_{cm} p_{cm}))]\}dz_s\,, \nonumber\\
   %%%%%%
   T^{ij}_{^3P_2}&=&\frac{1}{32\pi N_{ij}}\int_{-1}^1\frac{1}{20m_1m_2m_3m_4}\{F^{ij}_{(b)}[3 m_2 (m_4 (3 z_s^2-1) (-(p_{cm} p_{cm}^\prime-E_1 E_3 z_s))-E_1 m_3 z_s (z_s^2-1) (E_4+2 m_4))\nonumber\\
   &&+m_1 (m_3 (-3 (3 z_s^2-1) (p_{cm} p_{cm}^\prime-E_2 E_4 z_s)+2 m_4 z_s (m_2 (6 z_s^2-2)-3 E_2 (z_s^2-1))-6 E_4 m_2 z_s (z_s^2-1))\nonumber\\
   &&-3 E_3 m_4 z_s (z_s^2-1) (E_2+2 m_2))]+F^{ij}_{(c)}[3 m_2 (E_1 m_4 z_s (z_s^2-1) (E_3+2 m_3)-m_3 (3 z_s^2-1) (E_1 E_4 z_s+p_{cm} p_{cm}^\prime))\nonumber\\
   &&+m_1 (m_4 (-3 (3 z_s^2-1) (E_2 E_3 z_s+p_{cm} p_{cm}^\prime)+2 m_3 z_s (3 E_2 (z_s^2-1)+m_2 (2-6 z_s^2))+6 E_3 m_2 z_s (z_s^2-1))\nonumber\\
   &&+3 E_4 m_3 z_s (z_s^2-1) (E_2+2 m_2))]\}dz_s\,. \nonumber\\
\end{eqnarray}
One can estimate each partial wave amplitudes from Eq.~(\ref{eq:pw amp}). 
Inputting these partial wave amplitudes into Eq.~(\ref{eq:mod;T}), one can calculate the invariant mass spectrum and fit it to the data. 
Nevertheless, to get an impression about the partial wave scattering amplitudes intuitively, we list the analytical LO amplitudes as follows
\begin{eqnarray}\label{eq:LO}
    T_{^1S_0}^{11,LO}(s)& =& \frac{c_1(68m_1^4+8sm_1^2+5s^2)}{144\pi m_1^4}\,, \nonumber\\
    T_{^5S_2}^{11,LO} (s)& =& \frac{c_1(s+6m_1\sqrt{s}+14m_1^2)^2}{1800\pi m_1^4}\,, \nonumber\\
    T_{^3P_0}^{11,LO} (s)& =& 0\,, \nonumber\\
    T_{^3P_1}^{11,LO} (s)& =&-\frac{c_1(s-4m_1^2)}{12\pi m_1^2}\,, \nonumber\\
    T_{^3P_2}^{11,LO} (s)& =& 0\,, \nonumber\\
    %%%%%%%%%%%%%%%%%%%%%%%%%%%%%%%%%%%%%%%%%%%%%%%%%%%%%%%%%%%%%
    T_{^1S_0}^{12,LO}(s)& =& \frac{-10c_2 (m_1^3-m_2^2 m_1)^2-2 s^2 (3 m_2^2-8 m_2 m_1+m_1^2)+s (m_2^4+4 m_2^2 m_1^2+56 m_2 m_1^3+7 m_1^4)+5 s^3}{288 \sqrt{2} \pi  m_2 m_1^3 s}\,, \nonumber\\
    T_{^5S_2}^{12,LO} (s)& =&\frac{-c_2 (14 m_1^2+6 m_1 \sqrt{s}+s) ((m_1^2-m_2^2)^2-3 s^{3/2} (m_1+m_2)-14 m_1 m_2 s+3 \sqrt{s} (m_1-m_2)^2 (m_1+m_2)-s^2)}{3600 \sqrt{2} \pi m_1^3 m_2  s}\,, \nonumber\\
    T_{^3P_0}^{12,LO} (s)& =& 0\,, \nonumber\\
    T_{^3P_1}^{12,LO} (s)& =&\frac{c_2 (m_1+m_2)p_{cm}p_{cm}^\prime}{12\sqrt{2}\pi m_1^2 m_2}\,, \nonumber\\
    T_{^3P_2}^{12,LO} (s)& =& 0\,, \nonumber\\
    %%%%%%%%%%%%%%%%%%%%%%%%%%%%%%%%%%%%%%%%%%%%%%%%%%%%%%%%%%%%%
    T_{^1S_0}^{13,LO}(s)& =&\frac{c_5 (-10 (m_1^3-m_3^2 m_1)^2-2 s^2 (3 m_3^2-8 m_1 m_3+m_1^2)+s (m_3^4+4 m_1^2 m_3^2+56 m_1^3 m_3 +7 m_1^4)+5 s^3)}{288 \sqrt{2} \pi  m_1^3 m_3 s}\,, \nonumber\\
    T_{^5S_2}^{13,LO} (s)& =&\frac{-c_5 (14 m_1^2+6 m_1 \sqrt{s}+s) ((m_3^2-m_1^2)^2-3 s^{3/2} (m_1+m_3)-14 m_1 m_3 s+3 \sqrt{s} (m_1-m_3)^2 (m_1+m_3)-s^2)}{3600 \sqrt{2} \pi  m_1^3 m_3 s}\,, \nonumber\\
    T_{^3P_0}^{13,LO} (s)& =& 0\,, \nonumber\\
    T_{^3P_1}^{13,LO} (s)& =&\frac{c_5(m_1+m_3)*p_{cm}*p_{cm}^\prime}{12\sqrt{2}\pi m_1^2 m_3}\,, \nonumber\\
    T_{^3P_2}^{13,LO} (s)& =& 0\,, \nonumber\\
    %%%%%%%%%%%%%%%%%%%%%%%%%%%%%%%%%%%%%%%%%%%%%%%%%%%%%%%%%%%%%
    T_{^1S_0}^{22,LO}(s)& =& \frac{1}{288 \pi m_1^2  m_2^2 s^2}[-s^3 (3 m_2^2 (2 c_3+3 c_4)-4 m_1 m_2 (2 c_3+7 c_4)+3 m_1^2 (2 c_3+3 c_4))-s (2 c_3+c_4)*\nonumber\\
    && (3 m_2^2+4 m_1 m_2+3 m_1^2)(m_1^2-m_2^2)^2+(2 c_3+c_4) (m_1^2-m_2^2)^4+s^2 (8 c_3 (m_2^4+5  m_1^2m_2^2+m_1^4)\nonumber\\
    &&+c_4 (7 m_2^4-24  m_1m_2^3+74 m_1^2 m_2^2-24 m_1^3 m_2+7 m_1^4))+2 s^4 (c_3+2 c_4)]\,, \nonumber\\
    T_{^5S_2}^{22,LO} (s)& =&\frac{(2 c_3+c_4) (-(m_1^2-m_2^2)^2+3 s^{3/2} (m_1+m_2)+14 m_1 m_2 s-3 \sqrt{s} (m_1-m_2)^2 (m_1+m_2)+s^2)^2}{7200 \pi  m_1^2 m_2^2 s^2}\nonumber\\
    T_{^3P_0}^{22,LO} (s)& =& 0\,, \nonumber\\
    %T_{^3P_1}^{22,LO} (s)& =& -\frac{[c_3 (m_1^2+m_2^2)+2c_4m_1m_2 ]*s*\rho^2(s,m_1^2,m_2^2)}{48 m_1^2 m_2^2 }\,, \nonumber\\
    T_{^3P_1}^{22,LO} (s)& =& -\frac{[c_3 (m_1^2+m_2^2)+2c_4m_1m_2 ]p_{cm}^2}{12 m_1^2 m_2^2 }\,, \nonumber\\
    T_{^3P_2}^{22,LO} (s)& =& 0\,, \nonumber\\
    %%%%%%%%%%%%%%%%%%%%%%%%%%%%%%%%%%%%%%%%%%%%%%%%%%%%%%%%%%%%%
    T_{^1S_0}^{23,LO}(s)& =&\frac{1}{1152 \pi m_1^2  m_2 m_3 s^2}[-s^3 (3 (2 c_8+3 c_9) (2m_1^2+m_2^2+m_3^2)-4 m_1 (2 c_8+7 c_9) (m_2+m_3))+2 (2 c_8+c_9)* \nonumber\\
    &&(m_1^2-m_2^2)^2 (m_1^2-m_3^2)^2+s^2 (2 c_8 (m_2^4+4 m_1^2 (m_2^2+8 m_2 m_3+m_3^2)+6 m_2^2 m_3^2+m_3^4+8 m_1^4)+c_9 (m_2^4\nonumber\\
    &&+2 m_1^2 (5 m_2^2+64 m_2 m_3+5 m_3^2)+12 m_2^2 m_3^2-24 m_1^3 (m_2+m_3)-24 m_1 m_2 m_3 (m_2+m_3)+m_3^4+14 m_1^4))\nonumber\\
    &&-s (2 c_8+c_9) (4 m_1 m_2 m_3 (m_2^3+m_3^3)-3 m_1^4 (m_2^2+m_3^2)+3 m_2^2 m_3^2 (m_2^2+m_3^2)+3 m_1^2 (m_2^4-4 m_2^2 m_3^2+m_3^4)\nonumber\\
    &&+4 m_1^5 (m_2+m_3)-8 m_1^3 m_2 m_3 (m_2+m_3)+6 m_1^6)+4 s^4 (c_8+2 c_9)]\,, \nonumber\\
    T_{^5S_2}^{23,LO} (s)& =&\frac{1}{14400 \pi m_1^2  m_2 m_3 s^2}[(2 c_8+c_9) ((m_1^2-m_2^2)^2-3 s^{3/2} (m_1+m_2)-14 m_1 m_2 s+3 \sqrt{s} (m_1-m_2)^2*\nonumber\\
    && (m_1+m_2)-s^2)* ((m_1^2-m_3^2)^2-3 s^{3/2} (m_1+m_3)-14 m_1 m_3 s+3 \sqrt{s} (m_1-m_3)^2 (m_1+m_3)-s^2)]\nonumber\\
    T_{^3P_0}^{23,LO} (s)& =& 0\,, \nonumber\\
    T_{^3P_1}^{23,LO} (s)& =&\frac{[2c_8(m_1^2+m_2m_3)+c_9m_1(m_2+m_3)]p_{cm}p_{cm}^\prime}{48\pi m_1^2m_2m_3}\,, \nonumber\\
    T_{^3P_2}^{23,LO} (s)& =& 0\,, \nonumber\\
    %%%%%%%%%%%%%%%%%%%%%%%%%%%%%%%%%%%%%%%%%%%%%%%%%%%%%%%%%%%%%
    T_{^1S_0}^{33,LO}(s)& =& \frac{1}{288 \pi m_1^2 m_3^2  s^2}[-s^3 (3(m_1^2+ m_3^2) (2 c_6+3 c_7)-4 m_1 m_3 (2 c_6+7 c_7))-s (2 c_6+c_7) (3 m_3^2\nonumber\\
    &&+4 m_1 m_3+3 m_1^2)(m_1^2-m_3^2)^2+(2 c_6+c_7) (m_1^2-m_3^2)^4+s^2 (8 c_6 (m_3^4+5m_1^2 m_3^2 +m_1^4)\nonumber\\
    &&+c_7 (7 m_3^4-24 m_1 m_3^3+74 m_1^2 m_3^2-24 m_1^3 m_3+7 m_1^4))+2 s^4 (c_6+2 c_7)]\,, \nonumber\\
    T_{^5S_2}^{33,LO} (s)& =&\frac{(2c_6+c_7) (-(m_1^2-m_3^2)^2+3 s^{3/2} (m_1+m_3)+14 m_1 m_3 s-3 \sqrt{s} (m_1-m_3)^2 (m_1+m_3)+s^2)^2}{7200 \pi  m_1^2 m_3^2 s^2}\nonumber\\
    T_{^3P_0}^{33,LO} (s)& =& 0\,, \nonumber\\
    T_{^3P_1}^{33,LO} (s)& =& \frac{[c_6(m_1^2+m_3^2)+c_7m_1m_3]p_{cm}^{\prime 2}}{12\pi m_1^2m_3^2}\,, \nonumber\\
    T_{^3P_2}^{33,LO} (s)& =& 0\,, 
    %%%%%%%%%%%%%%%%%%%%%%%%%%%%%%%%%%%%%%%%%%%%%%%%%%%%%%%%%%%%%
\end{eqnarray}
One would find that the LO partial wave amplitudes are not simply the coupling constants, e.g., $c_i$, as the particles in the scattering have a spin $J=1$. Also, it shows the threshold behavior of the elastic and inelastic scattering partial waves. See $^3P_1$ waves for details.  

\section{The channels of $\chi_{cJ}\chi_{cJ}$}\label{app:chicJ}
\begin{figure}[htbp]
   \centering
   {\includegraphics[width=0.3\linewidth, height=0.1\textheight]{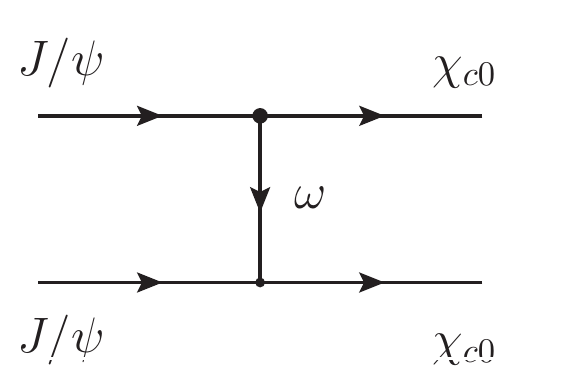}} 
     {\includegraphics[width=0.3\linewidth, height=0.1\textheight]{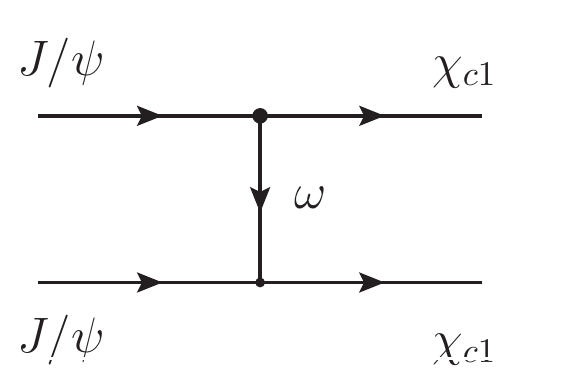}} 
    \caption{The $t$-channel meson exchange diagrams of $J/\psi J/\psi\to \chi_{c0}\chi_{c0}$ and $J/\psi J/\psi\to \chi_{c1}\chi_{c1}$. The $u$- channel ones are similar and not plotted here.  \label{fig:chicj}  }
\end{figure}

\begin{figure}[htbp]
   \centering
   {\includegraphics[width=0.9\linewidth, height=0.25\textheight]{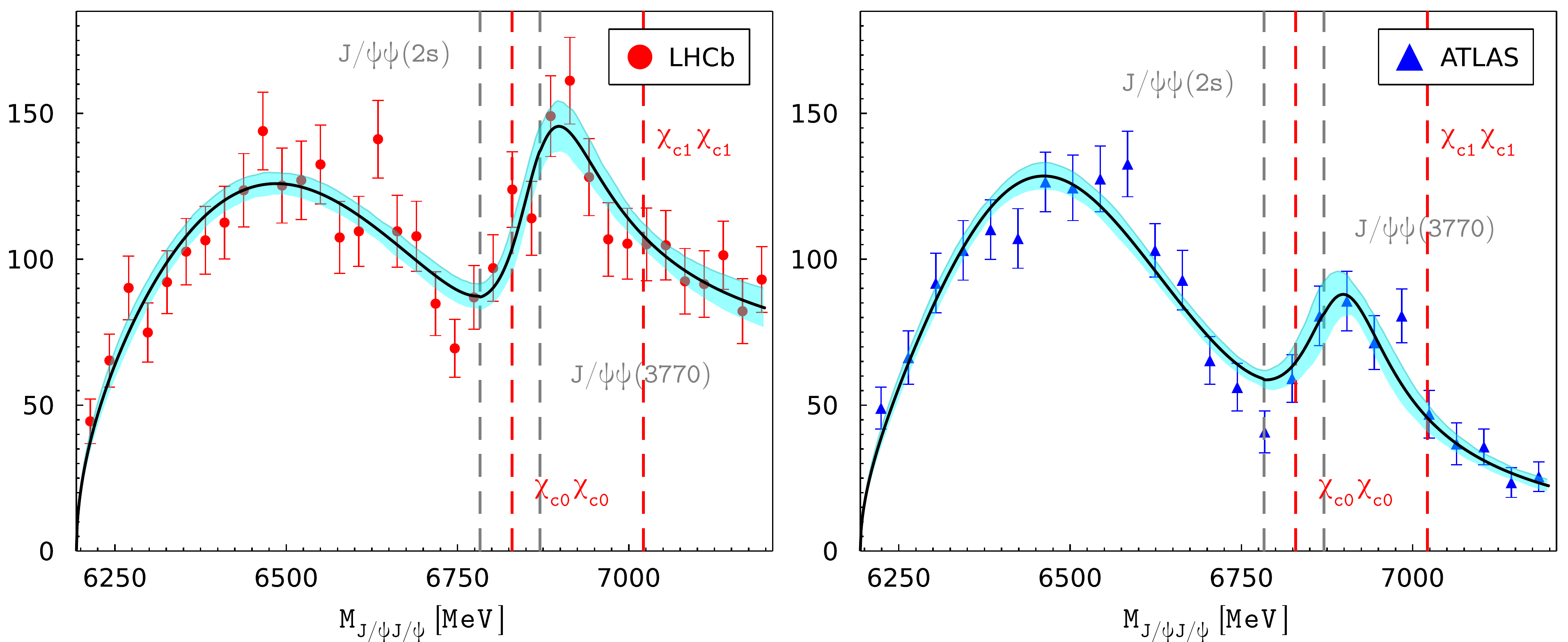} }
    \caption{The thresholds of $\chi_{c0}\chi_{c0}$, $\chi_{c1}\chi_{c1}$, $J/\psi \psi(2s)$, and $J/\psi \psi(3770)$.  \label{Fig:thresholds}  }
\end{figure}

About the intermediate channels of $\chi_{cJ}\chi_{cJ}$,  they contribute through the processes of $J/\psi J/\psi\to \chi_{c0}\chi_{c0} \to J/\psi J/\psi$ and $J/\psi J/\psi\to \chi_{c1}\chi_{c1} \to J/\psi J/\psi$, see Fig.\ref{fig:chicj}. 
Here the $\omega$ exchange dominates the diagrams.  
To elaborate it clearly, we construct the effective Lagrangians of $SVV$ and $AVV$ as 
\begin{eqnarray}
    \mathcal{L}_{SVV} &=& h_1\chi_{c0}\psi_\mu\langle V^\mu \rangle  \,, \nonumber\\
    \mathcal{L}_{AVV} &=& h_2\epsilon^{\mu\nu\alpha\beta}\psi_\mu\chi_{c1\nu}\langle V_{\alpha \beta} \rangle\,,
\end{eqnarray}
where $V_{\alpha \beta}=\nabla_\alpha V_\beta-\nabla_\beta V_\alpha$, and $V_\alpha$ is the Octec of lightest vector resonances \cite{Dai:2013joa}.
According to these effective Lagrangians, we can calculate the amplitudes listed in Fig.\ref{fig:chicj}. One has
\begin{eqnarray}
    T_{\chi_{c0}} &=& \frac{2h_1^2(\varepsilon_1\cdot\varepsilon_2)}{t-m_\omega^2}+\frac{2h_1^2(\varepsilon_1\cdot\varepsilon_2)}{u-m_\omega^2}\,, \nonumber\\
    T_{\chi_{c1}} &=& \frac{8h_2^2}{t-m_{\omega}^2}((  \varepsilon_3^*\cdot\varepsilon _4^*) ((  -p_3\cdot \varepsilon_1) (  p_4\cdot \varepsilon_2)-t (  \varepsilon_1 \cdot\varepsilon_2))+(  \varepsilon_2 \cdot\varepsilon_3^*) (t (  \varepsilon_1 \cdot\varepsilon_4^*)-(  -p_3\cdot \varepsilon_1) (  -p_2\cdot \varepsilon_4^*))\nonumber\\
    &&+(  p_1\cdot \varepsilon_3^*) (( -p_2\cdot \varepsilon_4^*) (  \varepsilon_1\cdot \varepsilon_2)-(  p_4\cdot\varepsilon_2) (  \varepsilon _1\cdot \varepsilon_4^*))) 
    +\frac{8h_2^2}{u-m_{\omega}^2} (( \varepsilon_3^*\cdot \varepsilon_4^*) ((-p_4\cdot   \varepsilon_1) ( p_3 \cdot \varepsilon_2)-u ( \varepsilon_1 \cdot \varepsilon_2))\nonumber\\
    &&+(\varepsilon_1 \cdot  \varepsilon_3^*) (u (  \varepsilon_2 \cdot\varepsilon_4^*)-( p_3\cdot  \varepsilon_2) ( p_1 \cdot \varepsilon_4^*))+( -p_2 \cdot \varepsilon_3^*) (( p_1 \cdot \varepsilon_4^*) ( \varepsilon_1 \cdot  \varepsilon_2)-( -p_4 \cdot \varepsilon_1) ( \varepsilon_2\cdot  \varepsilon_4^*)))\,.
\end{eqnarray}
As done in the previous sections, the partial wave decomposition of these amplitudes, e.g, S-waves are given as  
\begin{eqnarray}
    T_{^1S_0}^{\chi_{c0}} &=&     
    \frac{2\sqrt{3}}{3}T_{++}^{0,\chi_{c0}}-\frac{\sqrt{3}}{3}T_{00}^{0,\chi_{c0}}\notag\\
    &=&\frac{2h_1^2 (2 m_{J/\psi}^2+s) }{\sqrt{3} m_{J/\psi}^2 \sqrt{(s-4 m_{J/\psi}^2) (s-4 m_{\chi_{c0}}^2)}}\ln\left(\frac{2 (m_{J/\psi}^2+m_{\chi_{c0}}^2-m_\omega^2)-s+\sqrt{(s-4 m_{J/\psi}^2) (s-4 m_{\chi_{c0}}^2)}}{2 (m_{J/\psi}^2+m_{\chi_{c0}}^2-m_\omega^2)-s-\sqrt{(s-4 m_{J/\psi}^2) (s-4 m_{\chi_{c0}}^2)}}\right)\notag \,, \\
%%%%%%%%%%%%%%%%%%%%%%%%
    T_{^1S_0}^{\chi_{c1}} &=& \frac{1}{3}(2T_{++++}^{0,\chi_{c1}}+2T_{++--}^{0,\chi_{c1}}-2T_{++00}^{0,\chi_{c1}}-2T_{00++}^{0,\chi_{c1}}+T_{0000}^{0,\chi_{c1}})\notag\\
    &=&\frac{h_2^2 m_\omega^2 [4 m_{J/\psi}^2 m_{J\chi_{c1}}^2-s (m_{J/\psi}^2+m_{J\chi_{c1}}^2)] }{12 \pi  m_{J/\psi}^2 m_{J\chi_{c1}}^2 \sqrt{(s-4 m_{J/\psi}^2) (s-4 m_{J\chi_{c1}}^2)}}\ln \left(\frac{2 (m_{J/\psi}^2+m_{J\chi_{c1}}^2-m_\omega^2)-s+\sqrt{(s-4 m_{J/\psi}^2) (s-4 m_{J\chi_{c1}}^2)}}{2 (m_{J/\psi}^2+m_{J\chi_{c1}}^2-m_\omega^2)-s-\sqrt{(s-4 m_{J/\psi}^2) (s-4 m_{J\chi_{c1}}^2)}}\right)\notag\\
    &&+\frac{ h_2^2[4m_{J/\psi}^2 m_{J\chi_{c1}}^2-s (m_{J/\psi}^2+m_{J\chi_{c1}}^2)]}{12 \pi  m_{J/\psi}^2 m_{J\chi_{c1}}^2} \,.
\end{eqnarray}
From it, one sees very clearly the left hand cuts locate at ($-\infty, 5.908$~GeV) for $\chi_{c0}\chi_{c0}$  and  ($-\infty, 5.569$~GeV) for 
 $\chi_{c1}\chi_{c1}$. Obviously, These left hand cuts would be farther compared with the ones generated by the mesons exchange (such as $\pi\pi$, $\sigma$, $\eta$ and $\eta'$) of $J/\psi J/\psi \to J/\psi J/\psi$, $J/\psi \psi(2s)$, $J/\psi \psi(3770)$.  
Moreover, the thresholds of $\chi_{c0}\chi_{c0}$, $\chi_{c1}\chi_{c1}$ are 6.829~GeV and 7.021~GeV, respectively.  They are farther away from the structure (the peak around 6.90 GeV or the dip around 6.75GeV) of di-$J/\psi$ spectra, compared with that of $J/\psi \psi(2s)$ and $J/\psi \psi(3770)$. See Fig.\ref{Fig:thresholds}.
Hence, we ignore these two channels in our analysis.

\bibliographystyle{unsrt}
\bibliography{ref}

\end{document}